\documentclass[journal,draftclsnofoot,onecolumn,12pt,oneside]{IEEEtran} 

\usepackage{cite}

\ifCLASSINFOpdf
  \usepackage[pdftex]{graphicx}
  \usepackage{epstopdf} 
\else
\fi
\usepackage{amsmath}
\usepackage{algorithm}
\usepackage{algorithmic}

  \usepackage[caption=false,font=footnotesize]{subfig}
\usepackage{url}

\usepackage{amsthm}

\newtheoremstyle{mydefinition}
{}
{}
{}
{0pt}
{\bfseries}
{.}
{ }
{\thmname{#1}\thmnumber{ #2}: \thmnote{#3}}

\theoremstyle{mydefinition}

\newtheoremstyle{myremark}
{}
{}
{}
{0pt}
{\bfseries}
{.}
{ }
{\thmname{#1}\thmnumber{ #2}: \thmnote{#3}}
\theoremstyle{myremark}
\newcounter{counterchallenge}
\newtheorem{challenge}[counterchallenge]{Challenge}

\newtheoremstyle{remarkshort}
{}
{}
{}
{0pt}
{\bfseries}
{.}
{ }
{\thmname{#1}\thmnumber{ #2}}

\theoremstyle{remarkshort}

\newcounter{countertakeaway}
\theoremstyle{remarkshort}
\newtheorem{takeaway}[countertakeaway]{Takeaway}

\newcommand{\comment}[1]{{}}

\usepackage{amssymb}
\usepackage{xspace}
\usepackage{bm}
\usepackage{bbm}

\newcommand{\set}[1]{\ensuremath{\mathcal{#1}}\xspace} 
\newcommand{\mat}[1]{\ensuremath{\mathbf{#1}}\xspace} 
\renewcommand{\vec}[1]{\ensuremath{\mathbf{#1}}\xspace} 

\newcommand{\parens}[1]{\ensuremath{\left(#1\right)}\xspace}
\newcommand{\brackets}[1]{\ensuremath{\left[#1\right]}\xspace}
\newcommand{\braces}[1]{\ensuremath{\left\{#1\right\}}\xspace}
\newcommand{\bars}[1]{\ensuremath{\left\vert#1\right\vert}\xspace}

\renewcommand{\parens}[1]{{\left(#1\right)}\xspace}
\renewcommand{\brackets}[1]{{\left[#1\right]}\xspace}
\renewcommand{\braces}[1]{{\left\{#1\right\}}\xspace}
\renewcommand{\bars}[1]{{\left\vert#1\right\vert}\xspace}

\newcommand{\complex}{\ensuremath{\mathbb{C}}\xspace}

\newcommand{\floor}[1]{\ensuremath{\left\lfloor{#1}\right\rfloor}\xspace}

\newcommand{\setcomplex}{\ensuremath{\complex}}

\newcommand{\setmatrix}[3]{\ensuremath{#1^{#2 \times #3}}\xspace}

\newcommand{\setmatrixcomplex}[2]{\setmatrix{\setcomplex}{#1}{#2}}

\newcommand{\ctrans}{\ensuremath{^{{*}}}\xspace}

\newcommand{\entry}[2]{\ensuremath{\brackets{#1}_{#2}}\xspace}

\newcommand{\logtwo}[1]{\ensuremath{\mathrm{log}_{2}\parens{#1}}}

\DeclareMathOperator*{\argmax}{argmax}

\newcommand{\maxop}[1]{\ensuremath{\mathrm{max}\parens{#1}}\xspace}

\newcommand{\minop}[1]{\ensuremath{\mathrm{min}\parens{#1}}\xspace}

\newcommand{\st}{\ensuremath{\mathrm{s.t.~}}\xspace}
\newcommand{\opt}{\ensuremath{^{\star}}\xspace}

\newcommand{\powernoise}{\ensuremath{P_{\mathrm{noise}}}\xspace}

\newcommand{\powertx}{\ensuremath{P_{\mathrm{tx}}}\xspace}

\newcommand{\powertxue}{\ensuremath{\powertx^{\mathrm{UE}}}\xspace}
\newcommand{\powertxbs}{\ensuremath{\powertx^{\mathrm{BS}}}\xspace}

\newcommand{\snr}{\ensuremath{\mathsf{SNR}}\xspace}

\newcommand{\sinr}{\ensuremath{\mathsf{SINR}}\xspace}

\newcommand{\inr}{\ensuremath{\mathsf{INR}}\xspace}

\newcommand{\Nt}{\ensuremath{N_\mathrm{t}}\xspace} 
\newcommand{\Nr}{\ensuremath{N_\mathrm{r}}\xspace} 

\newcommand{\idx}[1]{\ensuremath{^{\parens{#1}}}\xspace}

\newcommand{\txdirsetmeas}{\set{T}_{\mathrm{tx}}\xspace}
\newcommand{\rxdirsetmeas}{\set{T}_{\mathrm{rx}}\xspace}

\newcommand{\txdirsetcb}{\ensuremath{\set{A}_{\mathrm{tx}}}\xspace}
\newcommand{\rxdirsetcb}{\ensuremath{\set{A}_{\mathrm{rx}}}\xspace}

\newcommand{\thetatx}{\ensuremath{\theta_{\mathrm{tx}}}\xspace}

\newcommand{\thetarx}{\ensuremath{\theta_{\mathrm{rx}}}\xspace}

\newcommand{\varthetatx}{\ensuremath{\vartheta_{\mathrm{tx}}}\xspace}

\newcommand{\varthetarx}{\ensuremath{\vartheta_{\mathrm{rx}}}\xspace}

\newcommand{\anglediff}[1]{\ensuremath{\measuredangle\parens{#1}}\xspace}

\newcommand{\Mtx}{\ensuremath{M_{\mathrm{tx}}}\xspace}
\newcommand{\Mrx}{\ensuremath{M_{\mathrm{rx}}}\xspace}

\newcommand{\Iij}{\ensuremath{\mathcal{I}_{ij}}\xspace}

\newcommand{\inrijmin}{\ensuremath{\inrij^{\mathrm{min}}}\xspace}
\newcommand{\inrijrng}{\ensuremath{\inrij^{\mathrm{rng}}}\xspace}

\newcommand{\labelue}{\mathrm{UE}}
\newcommand{\labeltx}{\mathrm{tx}}
\newcommand{\labelrx}{\mathrm{rx}}
\newcommand{\labelsum}{\mathrm{sum}}

\newcommand{\labeltgt}{\mathrm{tgt}}

\newcommand{\snrtx}{\ensuremath{\snr_{\labeltx}}\xspace}
\newcommand{\snrrx}{\ensuremath{\snr_{\labelrx}}\xspace}

\newcommand{\sinrtx}{\ensuremath{\sinr_{\labeltx}}\xspace}
\newcommand{\sinrrx}{\ensuremath{\sinr_{\labelrx}}\xspace}

\newcommand{\inrtx}{\ensuremath{\inr_{\labeltx}}\xspace}
\newcommand{\inrrx}{\ensuremath{\inr_{\labelrx}}\xspace}
\newcommand{\inrrxtgt}{\ensuremath{\inr_{\labelrx}^\labeltgt}\xspace}
\newcommand{\inrrxthresh}{\ensuremath{\inr_{\labelrx}^{\mathrm{tgt}}}\xspace}

\newcommand{\inrrxmin}{\ensuremath{\inr_{\labelrx}^{\mathrm{min}}}\xspace}

\newcommand{\setx}{\ensuremath{R_{\labeltx}}\xspace}
\newcommand{\serx}{\ensuremath{R_{\labelrx}}\xspace}
\newcommand{\sesum}{R_{\labelsum}\xspace}

\newcommand{\vhtx}{\vh_{\labeltx}\xspace}
\newcommand{\vhrx}{\vh_{\labelrx}\xspace}

\newcommand{\powernoiseue}{\ensuremath{\powernoise^{\labelue}}\xspace}
\newcommand{\powernoisebs}{\ensuremath{\powernoise^{\mathrm{BS}}}\xspace}

\newcommand{\setnbr}{\ensuremath{\mathcal{N}}\xspace}

\newcommand{\thtx}{\ensuremath{\theta_{\mathrm{tx}}}\xspace}

\newcommand{\thrx}{\ensuremath{\theta_{\mathrm{rx}}}\xspace}

\newcommand{\thtxi}{\ensuremath{\thtx\idx{i}}\xspace}

\newcommand{\thrxj}{\ensuremath{\thrx\idx{j}}\xspace}

\newcommand{\inrij}{\inr_{ij}}

\newcommand{\labelinit}{\mathrm{init}}

\newcommand{\pthtxrx}{\parens{\thetatx,\thetarx}\xspace}

\newcommand{\thtxrx}{\parens{\thetatx,\thetarx}\xspace}
\newcommand{\thtxopt}{\thtx\opt\xspace}
\newcommand{\thrxopt}{\thrx\opt\xspace}

\newcommand{\thtxinit}{{\theta}_{\labeltx}^{\labelinit}\xspace}
\newcommand{\thrxinit}{{\theta}_{\labelrx}^{\labelinit}\xspace}
\newcommand{\pthtx}{\parens{\thtx}\xspace}
\newcommand{\pthrx}{\parens{\thetarx}\xspace}

\newcommand{\pthtxrxinit}{\parens{\thtxinit,\thrxinit}\xspace}

\newcommand{\sCtx}{\sC_{\labeltx}\xspace}
\newcommand{\sCrx}{\sC_{\labelrx}\xspace}

\newcommand{\labelmax}{\mathrm{max}}

\newcommand{\sesumtgt}{\sesum^{\labeltgt}}
\newcommand{\sesummax}{\sesum^{\labelmax}}

\def\vf{{\vec{f}}}

\def\vh{{\vec{h}}}

\def\vw{{\vec{w}}}

\def\mH{{\mat{H}}}

\def\sC{{\set{C}}}

\usepackage{xspace}
\usepackage[acronym,nogroupskip,nonumberlist,nopostdot]{glossaries}
\makeglossaries

\newcommand{\steer}{\textsc{Steer}\xspace}
\newcommand{\steerp}{\textsc{Steer+}\xspace}

\usepackage{xcolor}

\newacronym{snr}{SNR}{signal-to-noise ratio}
\newacronym{sinr}{SINR}{signal-to-interference-plus-noise ratio}
\newacronym{inr}{INR}{interference-to-noise ratio}
\newacronym{sir}{SIR}{signal-to-interference ratio}
\newacronym{sqr}{SQR}{signal-to-quantization-noise ratio}
\newacronym{sqnr}{SQNR}{signal-to-quantization-plus-noise ratio}
\newacronym{ian}{IAN}{interference as noise}
\newacronym{ber}{BER}{bit error rate}
\newacronym{pn}{PN}{pseudorandom noise}
\newacronym{bfsk}{BFSK}{binary frequency shift keying}
\newacronym{fh}{FH}{frequency-hopped}
\newacronym{fh-bfsk}{FH-BFSK}{frequency-hopped binary frequency shift keying}
\newacronym{crc}{CRC}{cyclic redundancy check}
\newacronym{isi}{ISI}{intersymbol interference}
\newacronym{dsss}{DSSS}{direct-sequence spread spectrum}
\newacronym{ofdm}{OFDM}{orthogonal frequency-division multiplexing}
\newacronym{ofdma}{OFDMA}{orthogonal frequency-division multiple access}
\newacronym{sdr}{SDR}{software-defined radio}
\newacronym{tx}{TX}{transmitter}
\newacronym{rx}{RX}{receiver}
\newacronym{fdd}{FDD}{frequency-division duplexing}
\newacronym{tdd}{TDD}{time-division duplexing}
\newacronym{fdma}{FDMA}{frequency-division multiple access}
\newacronym{tdma}{TDMA}{time-division multiple access}
\newacronym{sdma}{SDMA}{space-division multiple access}
\newacronym[plural=MPCs]{mpc}{MPC}{multipath component}
\newacronym{mui}{MUI}{multi-user interference}
\newacronym{lsb}{LSB}{least significant bit}

\newacronym{qam}{QAM}{quadrature amplitude modulation}
\newacronym{mqam}{MQAM}{M-ary quadrature amplitude modulation}

\newacronym{ls}{LS}{least-squares}
\newacronym{lms}{LMS}{least mean squares}
\newacronym{rls}{RLS}{recursive least-squares}
\newacronym{rzf}{RZF}{regularized zero-forcing}
\newacronym{mmse}{MMSE}{minimum mean square error}
\newacronym{lmmse}{LMMSE}{linear minimum mean square error}
\newacronym{mse}{MSE}{mean square error}
\newacronym{fft}{FFT}{fast Fourier transform}
\newacronym{dft}{DFT}{discrete Fourier transform}
\newacronym{dtft}{DTFT}{discrete-time Fourier transform}
\newacronym{ctft}{CTFT}{continuous-time Fourier transform}
\newacronym{ml}{ML}{machine learning}
\newacronym[plural=NNs]{nn}{NN}{neural network}
\newacronym[plural=RNNs]{rnn}{RNN}{recurrent neural network}
\newacronym[plural=ADCs]{adc}{ADC}{analog-to-digital converter}
\newacronym[plural=DACs]{dac}{DAC}{digital-to-analog converter}
\newacronym[plural=FPGAs]{fpga}{FPGA}{field-programmable gate array}
\newacronym{evm}{EVM}{error vector magnitude}
\newacronym{enob}{ENOB}{effective number of bits}
\newacronym{zf}{ZF}{zero-forcing}
\newacronym{rv}{r.v.}{random variable}
\newacronym{omp}{OMP}{orthogonal matching pursuit}
\newacronym{svd}{SVD}{singular value decomposition}

\newacronym{sdp}{SDP}{semidefinite programming}
\newacronym{psd}{PSD}{positive semidefinite}
\newacronym{nsd}{NSD}{negative semidefinite}

\newacronym{ks}{K-S}{Kolmogorov-Smirnov}

\newacronym{mad}{MAD}{median absolute deviation around the median}

\newacronym{agc}{AGC}{automatic gain control}
\newacronym{rf}{RF}{radio frequency}
\newacronym{if}{IF}{intermediate frequency}
\newacronym{los}{LOS}{line-of-sight}
\newacronym{nlos}{NLOS}{non-line-of-sight}
\newacronym{ple}{PLE}{path loss exponent}
\newacronym[plural=dB,firstplural=decibels (dB)]{db}{dB}{decibel}
\newacronym[plural=dBm,firstplural=decibel milliwatts (dBm)]{dbm}{dBm}{decibel milliwatts}
\newacronym{pa}{PA}{power amplifier}
\newacronym{lna}{LNA}{low noise amplifier}
\newacronym{vga}{VGA}{variable gain amplifier}
\newacronym{cw}{CW}{continuous wave}
\newacronym{papr}{PAPR}{peak-to-average power ratio}
\newacronym{usrp}{USRP}{Universal Software Radio Peripheral}
\newacronym{irr}{IRR}{image rejection ratio}
\newacronym{lo}{LO}{local oscillator}
\newacronym{vm}{VM}{vector modulator}
\newacronym{mmwave}{mmWave}{millimeter wave}
\newacronym{eirp}{EIRP}{effective isotropic radiated power}
\newacronym{rsrp}{RSRP}{reference signal received power}

\newacronym{csma}{CSMA}{carrier-sense multiple access}
\newacronym{csmaca}{CSMA/CA}{carrier-sense multiple access with collision avoidance}
\newacronym{csmacd}{CSMA/CD}{carrier-sense multiple access with collision detection}
\newacronym{mac}{MAC}{medium access control}
\newacronym{phy}{PHY}{physical layer}
\newacronym{4g}{4G}{fourth generation}
\newacronym{lte}{LTE}{Long-Term Evolution}
\newacronym{4glte}{4G LTE}{\gls{4g} \gls{lte}}
\newacronym{5g}{5G}{fifth generation}
\newacronym{nr}{NR}{New Radio}
\newacronym{5gnr}{5G NR}{5G New Radio}
\newacronym{ieee}{IEEE}{Institute of Electrical and Electronics Engineers}
\newacronym{wifi}{Wi-Fi}{IEEE 802.11}
\newacronym{lan}{LAN}{local area network}
\newacronym{wlan}{WLAN}{wireless local area network}
\newacronym[plural=BSs]{bs}{BS}{base station}
\newacronym[plural=SBSs]{sbs}{SBS}{small-cell base station}
\newacronym[plural=FD-SBSs]{fdsbs}{FD-SBS}{\gls{fd}-enabled \gls{sbs}}
\newacronym[plural=MBSs]{mbs}{MBS}{macrocell base station}
\newacronym[plural=UEs]{ue}{UE}{user equipment}
\newacronym{ul}{UL}{uplink}
\newacronym{dl}{DL}{downlink}
\newacronym{qos}{QoS}{Quality of Service}
\newacronym{fcc}{FCC}{Federal Communications Commission}
\newacronym{iab}{IAB}{integrated access and backhaul}
\newacronym{fab}{FAB}{fixed access and backhaul}
\newacronym{hetnet}{HetNet}{heterogeneous network}

\newacronym{siso}{SISO}{single-input single-output}
\newacronym{mimo}{MIMO}{multiple-input multiple-output}
\newacronym{sumimo}{SU-MIMO}{single-user \gls{mimo}}
\newacronym{mumimo}{MU-MIMO}{multi-user \gls{mimo}}
\newacronym{bf}{BF}{beamforming}
\newacronym{ca}{CA}{constant amplitude}
\newacronym{ula}{ULA}{uniform linear array}
\newacronym{upa}{UPA}{uniform planar array}
\newacronym[\glslongpluralkey={angles of arrival}]{aoa}{AoA}{angle of arrival}
\newacronym[\glslongpluralkey={angles of departure}]{aod}{AoD}{angle of departure}
\newacronym{dof}{DoF}{degrees of freedom}
\newacronym{csi}{CSI}{channel state information}
\newacronym{csit}{CSIT}{\gls{csi} at the transmitter}
\newacronym{csir}{CSIR}{\gls{csi} at the receiver}
\newacronym{cs}{CS}{compressed sensing}

\newacronym{fd}{FD}{in-band full-duplex}
\newacronym{hd}{HD}{half-duplex}
\newacronym{si}{SI}{self-interference}
\newacronym{sic}{SIC}{self-interference cancellation}
\newacronym{soi}{SoI}{signal of interest}
\newacronym{asic}{A-SIC}{analog \acrlong{sic}}
\newacronym{dsic}{D-SIC}{digital \gls{sic}}
\newacronym{star}{STAR}{simultaneous transmit and receive}
\newacronym{warp}{WARP}{Wireless Open-Access Research Platform}
\newacronym{bfc}{BFC}{beamforming cancellation}
\newacronym{ipi}{IPI}{inter-panel-interference}
\newacronym{ipic}{IPIC}{inter-panel-interference cancellation}

\newacronym{qcqp}{QCQP}{quadratically-constrained quadratic programming}
\newacronym{pdf}{PDF}{probability density function}
\newacronym{cdf}{CDF}{cumulative density function}
\newacronym{iid}{i.i.d.}{independently and identically distributed}

\newacronym{elf}{ELF}{extremely low frequency}
\newacronym{slf}{SLF}{super low frequency}
\newacronym{ulf}{ULF}{ultra low frequency}
\newacronym{vlf}{VLF}{very low frequency}
\newacronym{lf}{LF}{low frequency}
\newacronym{mf}{MF}{medium frequency}
\newacronym{hf}{HF}{high frequency}
\newacronym{vhf}{VHF}{very high frequency}
\newacronym{uhf}{UHF}{ultra high frequency}
\newacronym{shf}{SHF}{super high frequency}
\newacronym{ehf}{EHF}{extremely high frequency}
\newacronym{thf}{THF}{tremendously high frequency}

\newacronym{wncg}{WNCG}{Wireless Networking and Communications Group}
\newacronym{linc}{LINC}{Laboratory of Informatics, Networks, and Communications}
\newacronym{ut}{UT Austin}{The University of Texas at Austin}
\newacronym{uiuc}{UIUC}{University of Illinois at Urbana-Champaign}
\newacronym{usc}{USC}{University of Southern California}
\newacronym{mit}{MIT}{Massachusetts Institute of Technology}
\newacronym{berkeley}{UC Berkeley}{University of California, Berkeley}
\newacronym{osu}{OSU}{Ohio State University}

\newcommand{\ula}{\gls{ula}\xspace}
\newcommand{\ulas}{\glspl{ula}\xspace}

\newcommand{\mmwave}{\gls{mmwave}\xspace}
\newcommand{\mimo}{\gls{mimo}\xspace}

\newcommand{\ue}{\gls{ue}\xspace}
\newcommand{\ues}{\glspl{ue}\xspace}
\newcommand{\rf}{\gls{rf}\xspace}

\newcommand{\gcdf}{\gls{cdf}\xspace}
\newcommand{\gpcdf}{\glspl{cdf}\xspace}
\newcommand{\bs}{\gls{bs}\xspace}
\newcommand{\bss}{\glspl{bs}\xspace}
\newcommand{\gsnr}{\gls{snr}\xspace}
\newcommand{\ginr}{\gls{inr}\xspace}
\newcommand{\gsinr}{\gls{sinr}\xspace}

\newcommand{\gpsnr}{\glspl{snr}\xspace}
\newcommand{\gpinr}{\glspl{inr}\xspace}
\newcommand{\gpsinr}{\glspl{sinr}\xspace}

\newcommand{\tdd}{\gls{tdd}\xspace}

\newcommand{\secref}[1]{Section~\ref{#1}}

\newcommand{\figref}[1]{\figurename~\ref{#1}}
\newcommand{\algref}[1]{Algorithm~\ref{#1}}

\usepackage{mathtools}

\begin{document}

%
\title{Real-World Evaluation of Full-Duplex Millimeter Wave Systems using\\Off-the-Shelf 60 GHz Radios}
\title{Real-World Evaluation of Full-Duplex\\Millimeter Wave Communication Systems}

%
%
%

\author{%
    Ian~P.~Roberts,~%
    Yu Zhang,~%
    Tawfik Osman,~%
    and Ahmed Alkhateeb%
    \thanks{I.~P.~Roberts is with the Department of Electrical and Computer Engineering at UCLA. Y.~Zhang, T.~Osman, and A.~Alkhateeb are with the Wireless Intelligence Lab at Arizona State University. Corresponding author: I.~P.~Roberts (ianroberts@ucla.edu).}
}

\maketitle

\begin{abstract}
Noteworthy strides continue to be made in the development of full-duplex \mmwave communication systems, but most of this progress has been built on theoretical models and validated through simulation.
In this work, we conduct a long overdue real-world evaluation of full-duplex \mmwave systems using off-the-shelf 60 GHz phased arrays.
Using an experimental full-duplex base station, we collect over 200,000 measurements of self-interference by electronically sweeping its transmit and receive beams across a dense spatial profile, shedding light on the effects of the environment, array positioning, and beam steering direction.
We then call attention to five key challenges faced by practical full-duplex \mmwave systems and, with these in mind, propose a general framework for beamforming-based full-duplex solutions.
Guided by this framework, we introduce a novel solution called \steerp, a more robust version of recent work called \steer, and experimentally evaluate both in a real-world setting with actual downlink and uplink users.
Rather than purely minimize self-interference as with \steer, \steerp makes use of additional measurements to maximize spectral efficiency, which proves to make it much less sensitive to one's choice of design parameters. 
We experimentally show that \steerp can reliably reduce self-interference to near or below the noise floor while maintaining high SNR on the downlink and uplink, thus enabling full-duplex operation purely via beamforming.

\end{abstract}

\glsresetall

\section{Introduction} \label{sec:introduction}





There is exciting potential for a \mmwave transceiver to transmit and receive in a full-duplex fashion by crafting its highly directional beams such that they couple low self-interference \cite{xia_2017,li_2014,roberts_wcm,liu_beamforming_2016}.
Recent work (e.g., \cite{liu_beamforming_2016,satyanarayana_hybrid_2019,cai_robust_2019,zhu_uav_joint_2020,roberts_lonestar,da_silva_2020,lopez_analog_2022,roberts_steer}) has shown that strategic beamforming alone can in fact reduce self-interference to levels sufficiently low for full-duplex \mmwave operation, rendering additional analog or digital cancellation potentially unnecessary.
Upgrading \mmwave \bss with full-duplex capability could make better use of radio resources to deliver higher throughput and reduced latency---especially in multi-hop, wirelessly-backhauled networks---even when users remain as half-duplex devices \cite{ozgur_isit_2017,gupta_fdiab}.
For example, a sectorized \mmwave \bs with inter-sector full-duplex capability could independently and thus more flexibly schedule its sectors to more optimally meet uplink and downlink demands and to juggle access and wireless backhauling \cite{gupta_fdiab,3GPP_IAB_2,suk_iab_2022,roberts_chapter}.
Applications of full-duplex can also be found in joint communication and sensing \cite{jcas_wcm_2021,xiao_jsac_2022}, spectrum sharing \cite{liao_mag_2015}, and secure communications \cite{wang_coml_2022}.





The majority of existing work developing beamforming-based solutions for full-duplex \mmwave systems has been built on theory with impractical assumptions and evaluated through simulation using idealized models.
Most commonly in the literature, \mmwave self-interference has been modeled using the spherical-wave channel \cite{spherical_2005} to capture idealized near-field propagation between the arrays of a full-duplex transceiver, often in conjunction with a ray-based model to incorporate reflections off the environment \cite{li_2014,satyanarayana_hybrid_2019,rajagopal_2014,lee_2015}.
These channel models have been used to construct and evaluate beamforming designs for full-duplex systems, but their real-world validity has not yet been confirmed.
A recent measurement campaign \cite{roberts_att_angular} and follow-on modeling \cite{roberts_att_modeling} provides perhaps the most comprehensive characterization of \mmwave self-interference to date.
Although \cite{roberts_att_angular,roberts_att_modeling} do not propose a particular \mimo channel model for self-interference, they do show that their measurements do not align with the spherical-wave model \cite{spherical_2005}.
Works \cite{roberts_att_angular,roberts_att_modeling} were both based on measurements taken in an anechoic chamber with the same 28 GHz phased array platform. 
The research community would benefit from more measurements of self-interference collected with a variety of phased array platforms in several orientations, at multiple carrier frequencies, and in real-world environments beyond an anechoic chamber.

In addition to relying on idealized self-interference channel modeling for evaluation, existing beamforming-based full-duplex solutions also have a number of other practical shortcomings. 
Real-time estimation of the self-interference, downlink, and uplink \mimo channels is not viable in today's real-world \mmwave systems but has been assumed in most work.
In practice, such as in 5G and IEEE 802.11ay, \mmwave systems circumvent high-dimensional \mimo channel estimation and instead rely on \textit{beam alignment} measurements to identify beams that close the link between a \bs and \ue \cite{heath_overview_2016,ethan_beam,junyi_wang_beam_2009,aa_initial_2017}.
Many existing designs also do not account for the digitally-controlled nature of analog beamforming networks, raising questions on how these designs will fare when implemented on actual phased arrays. 
Recent work \cite{roberts_steer} addresses many of these practical considerations through the design of \steer, a beam selection methodology for full-duplex \mmwave systems, inspired by the small-scale spatial variability of self-interference discovered in \cite{roberts_att_angular}.
In \cite{roberts_steer}, \steer was evaluated through measurements of self-interference taken in an anechoic chamber with the same 28 GHz platform as \cite{roberts_att_angular,roberts_att_modeling}, which yielded promising results. 
It is unclear, however, what level of success \steer would see when implemented on other phased arrays, in real-world environments, and when downlink/uplink performance is measured rather than simulated.

In this work, we investigate real-world full-duplex \mmwave systems through measurements of self-interference and through experimental evaluation of full-duplex solutions.
The principal contributions of the work herein are summarized as follows.
\begin{itemize}
    \item \textbf{Measurements of Self-Interference (\secref{sec:measurements}).} 
    We use an experimental full-duplex \bs---comprised of off-the-shelf 60 GHz phased arrays---to collect over 200,000 measurements of self-interference in three different environments and in a variety of system configurations. 
    By densely sweeping our transmit and receive beams with fine resolution, we inspect the spatial profile of self-interference and investigate the impacts of the environment, array positioning, and beam steering direction. 
    \item \textbf{Practical Challenges and Framework (\secref{sec:challenges}).}
    We then call attention to five key challenges faced by \textit{real-world} full-duplex \mmwave systems.
    With these challenges in mind, we present a general framework for beamforming-based full-duplex solutions that can steer the direction of future research.
    Core to our framework is that it is measurement-driven in the sense that it does not rely on estimation of high-dimensional \mimo channels (which is practically difficult) but rather measurements over such channels for particular beams.
    \item \textbf{Experimental Evaluation of Full-Duplex Solutions (\secref{sec:steer}).}
    Using our 60 GHz phased arrays, we then conduct a real-world evaluation of existing work \steer \cite{roberts_steer} by distributing \ues around our experimental full-duplex \bs in an indoor environment.
    Our measurements show that \steer can indeed be a fruitful full-duplex solution but that it can be sensitive to one's choice of design parameters.
    Motivated by this, we improve upon \steer to create \steerp, a more robust solution that experimentally proves to offer higher \gpsinr and spectral efficiencies than \steer.
    To our knowledge, these experimental evaluations of \steer and \steerp are the first real-world evaluations of their kind, complete with measurements of self-interference, downlink and uplink \gpsnr, and cross-link interference.
\end{itemize}



\section{Experimental Full-Duplex mmWave System} \label{sec:setup}


\begin{figure*}
    \centering
    \subfloat[60 GHz phased arrays.]{\includegraphics[width=\linewidth,height=0.215\textheight,keepaspectratio]{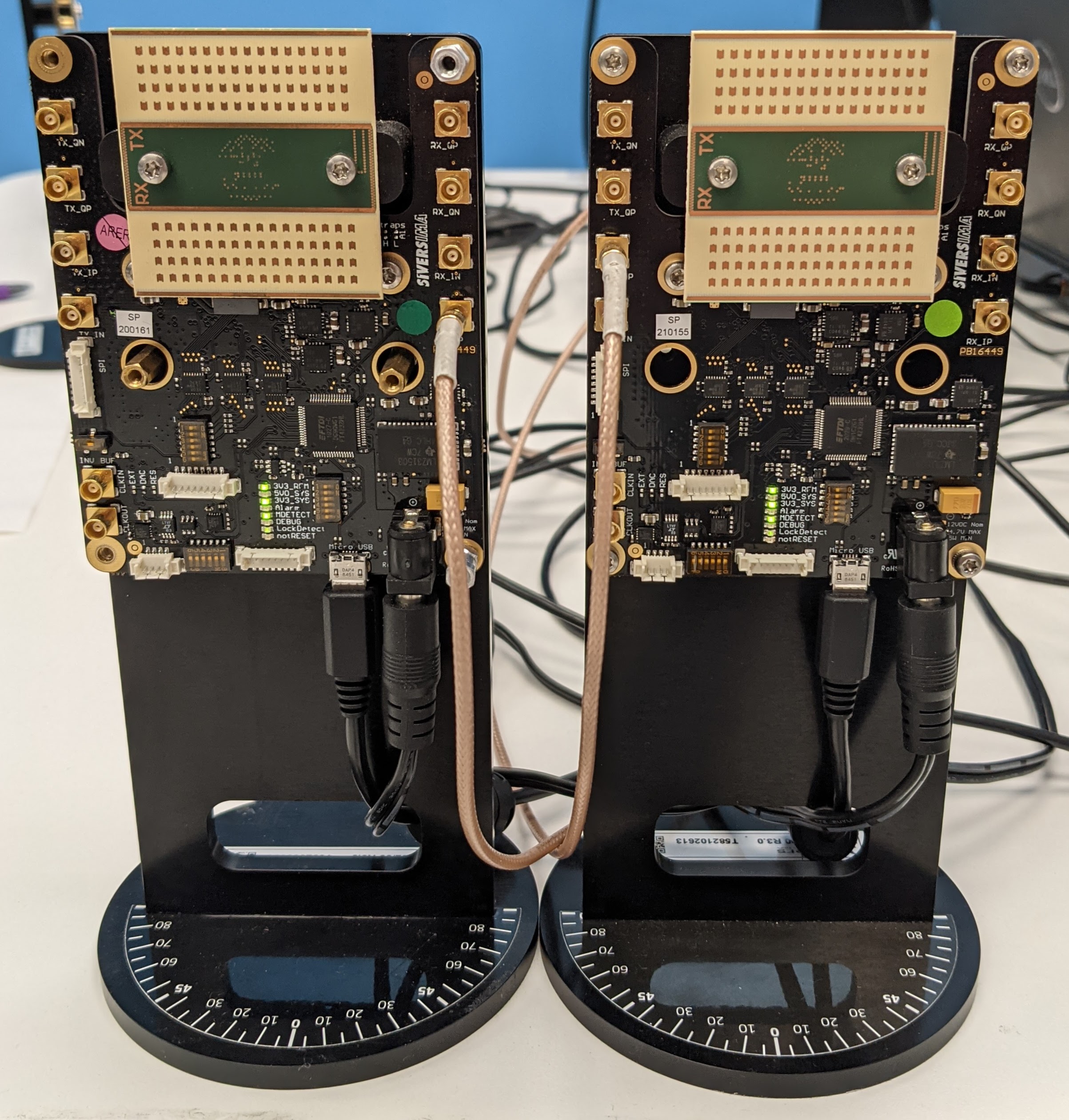}
        \label{fig:arrays}}
    \hfill
    \subfloat[Lab environment.]{\includegraphics[width=\linewidth,height=0.215\textheight,keepaspectratio]{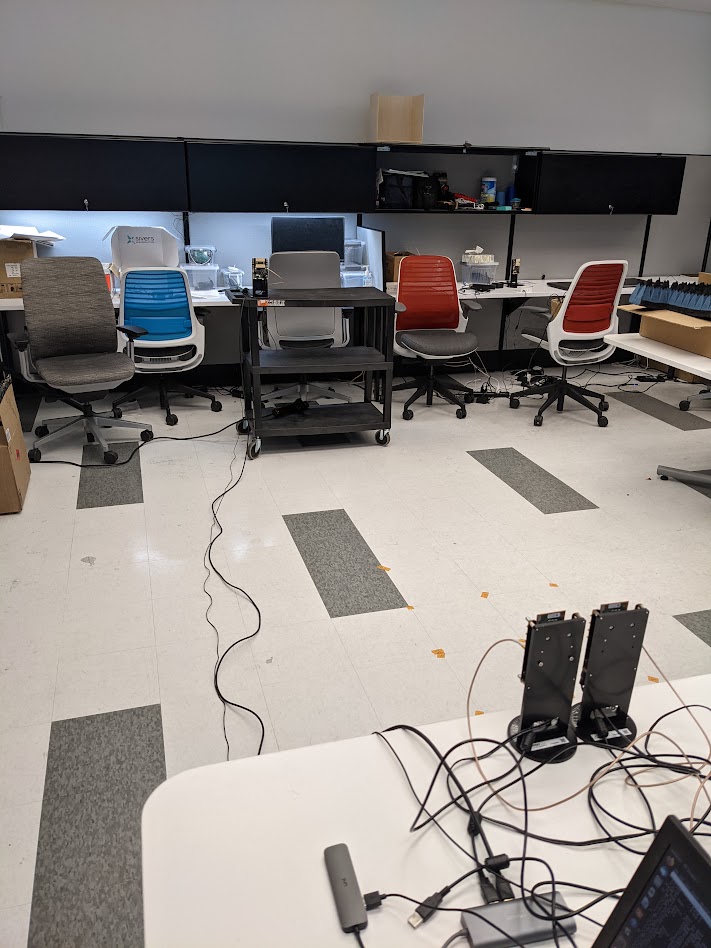}
        \label{fig:lab}}
    \hfill
    \subfloat[Lobby environment.]{\includegraphics[width=\linewidth,height=0.215\textheight,keepaspectratio]{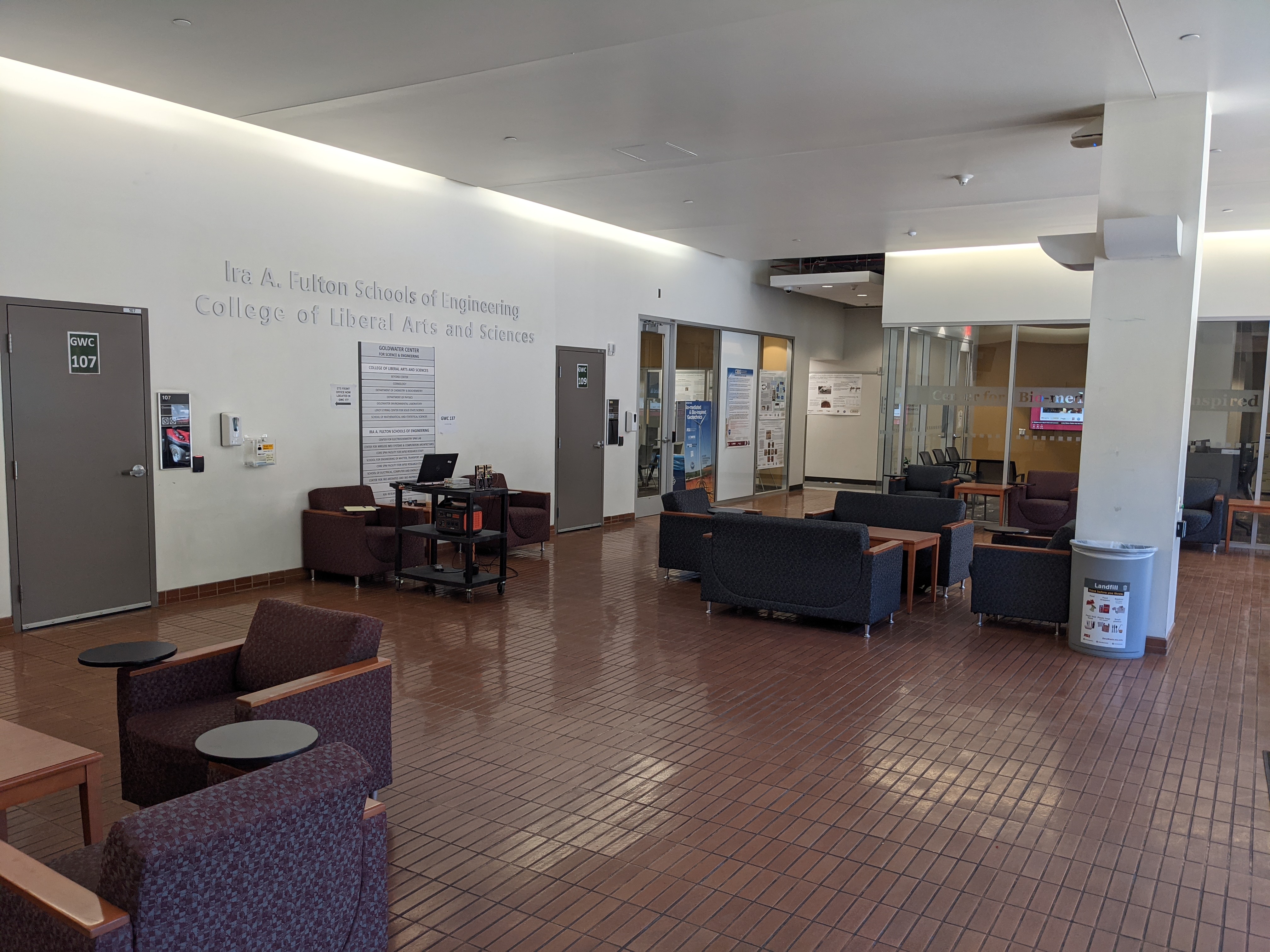}
        \label{fig:lobby}}
    \caption{(a) The 60 GHz phased arrays used throughout this work. By transmitting with one and receiving with the other in-band, we measure self-interference in a variety of settings. The (b) lab environment and (c) lobby environment in which measurements were collected. Direct coupling between the arrays along with reflections off the environment contribute to self-interference.}
    \label{fig:settings}
\end{figure*}

Consider a \mmwave communication system where a full-duplex \bs transmits downlink to one half-duplex \ue while simultaneously receiving uplink from another half-duplex \ue in-band; an illustration of this can be seen in \figref{fig:steer-setup} in \secref{sec:steer}.
Serving these two \ues in a full-duplex fashion manifests self-interference at the \bs and cross-link interference at the downlink \ue.
We assume the \bs employs separate phased arrays for transmission and reception, whereas each \ue is equipped with a single omnidirectional antenna, but this is not a necessary assumption.

\begin{figure*}
    \centering
    \subfloat[Measured beam pattern.]{\includegraphics[width=\linewidth,height=0.17\textheight,keepaspectratio]{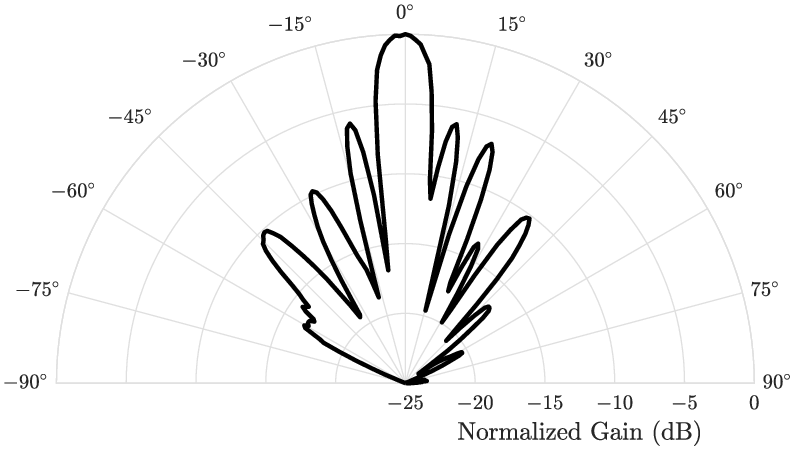}
        \label{fig:beam}}
    \quad
    \subfloat[Beam sweeping measurements.]{\includegraphics[width=\linewidth,height=0.16\textheight,keepaspectratio]{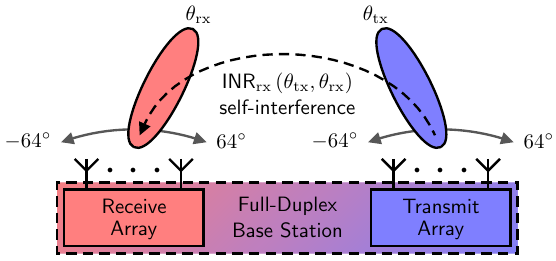}
        \label{fig:sweep}}
    \caption{(a) The measured pattern of a beam steered broadside by one of our 60 GHz phased arrays. (b) Measurements of self-interference were collected by sweeping the transmit and receive beams each in azimuth from $-64^\circ$ to $64^\circ$ in $1^\circ$ steps.}
    \label{fig:beam-sweep}
    \vspace{-0.5cm}
\end{figure*}

Our measurements herein were collected using off-the-shelf 60 GHz phased arrays, two of which are shown side-by-side in \figref{fig:arrays}.
One phased array served as a transmitter while an identical one acted as a receiver, together comprising an experimental full-duplex \bs.
Each phased array is equipped with separate, vertically-stacked transmit and receive arrays, each being a 16-element horizontal \ula with half-wavelength spacing and having four vertical elements for static beamforming.
The transmit array can be electronically steered to produce a beam in some relative azimuth direction $\thetatx$ via digitally-controlled analog beamforming weights $\vf\pthtx \in \setcomplex^{\Nt}$, where $\Nt = 16$ antennas.
Likewise, the receive array can be steered toward $\thetarx$ via weights $\vw\pthrx \in \setcomplex^{\Nr}$, where $\Nr = 16$.
The measured gain pattern of a beam steered broadside by one of our phased arrays is shown in \figref{fig:beam}.
The half-power beamwidth is approximately $10^\circ$, and side lobes are at least $6$ dB below the main lobe.


When the \bs serves a downlink \ue and an uplink \ue in a full-duplex fashion, the \gpsinr of the downlink and uplink can be expressed as
\begin{align}
\sinrtx\pthtx = \frac{\snrtx\pthtx}{1+\inrtx}, \qquad
\sinrrx\pthtxrx = \frac{\snrrx\pthrx}{1+\inrrx\pthtxrx}.
\end{align}
Here, $\snrtx$ and $\snrrx$ are the downlink and uplink \gpsnr, while $\inrtx$ and $\inrrx$ are their \gpinr due to cross-link interference and self-interference, respectively.
These \gpsnr are respectively functions of the transmit and receive beams as 
\begin{align}
\snrtx\pthtx &= \frac{\powertxbs \cdot \bars{\vhtx\ctrans \vf\pthtx}^2}{\powernoiseue}, \qquad 
\snrrx\pthrx = \frac{\powertxue \cdot \bars{\vw\pthrx\ctrans \vhrx}^2}{\powernoisebs}, \label{eq:snr-trx}
\end{align}
where $\powertxbs$ and $\powertxue$ are the \bs and uplink \ue transmit powers; $\powernoisebs$ and $\powernoiseue$ are the \bs and downlink \ue noise powers; and $\vhtx\in\setcomplex^{\Nt}$ and $\vhrx\in\setcomplex^{\Nr}$ are the downlink and uplink channels.
When serving downlink and uplink in a full-duplex fashion, self-interference manifests between the transmit and receive arrays of the full-duplex \bs.
Assuming a linear model, the \ginr of the resulting self-interference is a function of the transmit and receive beams and is written as
\begin{align}
\inrrx\pthtxrx = \frac{\powertxbs \cdot \bars{\vw\pthrx\ctrans \mH \vf\pthtx}^2}{\powernoisebs},
\end{align}
where $\bars{\vw\pthrx\ctrans \mH \vf\pthtx}^2$ captures the transmit and receive beam coupling over the self-interference channel $\mH \in \setmatrixcomplex{\Nr}{\Nt}$.
Cross-link interference $\inrtx$ can be expressed as
\begin{align}
\inrtx = \frac{\powertxue \cdot \bars{h}^2}{\powernoiseue},
\end{align}
which only depends on its channel $h \in \setcomplex$, since the \ues are assumed as single-antenna devices.

\section{Self-Interference Measurements} \label{sec:measurements}

Unless specified otherwise, the measurements throughout this section were collected with the transmit and receive arrays positioned next to one another side-by-side, as depicted in \figref{fig:arrays} and \figref{fig:sweep}, with the array centers separated by 10 cm.
Looking out from the arrays' perspective, the transmit array was on the right and the receive array on the left.
As denoted in \figref{fig:sweep}, broadside corresponds to an azimuth of $0^\circ$, steering rightward is an increase in azimuth, and steering leftward is a decrease in azimuth.
Using our experimental full-duplex \bs, we collected measurements of self-interference by sweeping the transmit and receive beams over respective spatial profiles $\txdirsetcb$ and $\rxdirsetcb$ across a bandwidth of 100 MHz at a carrier frequency of 60 GHz. 
\begin{align}
\txdirsetcb &= \braces{\thetatx\idx{i} : i = 1, \dots, \Mtx}, \qquad 
\rxdirsetcb = \braces{\thetarx\idx{j} : j = 1, \dots, \Mrx}
\end{align}
For each transmit-receive steering combination $\pthtxrx$, self-interference $\inrrx\pthtxrx$ was measured and recorded to form a set of $\Mtx \times \Mrx$ measurements, written as
\begin{align}
\braces{\inrrx\pthtxrx : \thtx \in \txdirsetcb, \thrx \in \rxdirsetcb}.
\end{align}
Throughout this section, we swept our transmit and receive beams over identical spatial profiles from $-64^\circ$ to $64^\circ$ in $1^\circ$ steps, meaning $\txdirsetcb = \rxdirsetcb = \braces{-64^\circ,-63^\circ,\dots,64^\circ}$, for a total of $129\times129 = 16,641$ measurements collected per sweep.
Note that the level of self-interference we are interested in measuring is purely that coupled by the transmit and receive beams and does not include any forms of self-interference cancellation.
The collected measurements were post-processed in various ways, whose results are presented in the subsections that follow.
By collecting measurements in a variety of environments and system configurations, we gained insights on the strength and spatial characteristics of self-interference, as well as factors that impact such.
In particular, we look at the effects of the following on self-interference in this section:
(i) the environment;
(ii) horizontal separation of the transmit and receive arrays;
(iii) angular separation of the transmit and receive arrays;
(iv) swapping the positioning of the transmit and receive arrays; and 
(v) slightly shifting the steering directions of the transmit and receive beams (on the order of one degree). 

\subsection{Self-Interference in Various Environments}
Prior work measuring \mmwave self-interference is quite limited for the most part, and the most extensive measurements thus far \cite{roberts_att_angular,roberts_att_modeling} were collected exclusively in an anechoic chamber.
In this work, to better understand self-interference in real-world settings, we collected measurements in three different environments: 
\begin{itemize}
    \item \textbf{Anechoic chamber.} The arrays were placed in an anechoic chamber without any significant reflectors. In principle, this allows us to solely inspect the \textit{direct} coupling between the arrays.
    \item \textbf{Lab (or office-like) environment.} Shown in \figref{fig:lab}, the lab environment contains desks, tables, chairs, metal cabinets, and miscellaneous furniture and equipment. The arrays were positioned on the edge of a table, facing outward into the room.
    \item \textbf{Lobby environment.} Shown in \figref{fig:lobby}, the lobby of the Goldwater Center at Arizona State University contains sofas, chairs, wooden coffee tables, tile flooring, windows, and support columns. The arrays were placed on the edge of a plastic cart, facing outward into the room.
\end{itemize}

\begin{figure*}
    \centering
    \subfloat[Anechoic chamber.]{\includegraphics[width=0.32\linewidth,height=\textheight,keepaspectratio]{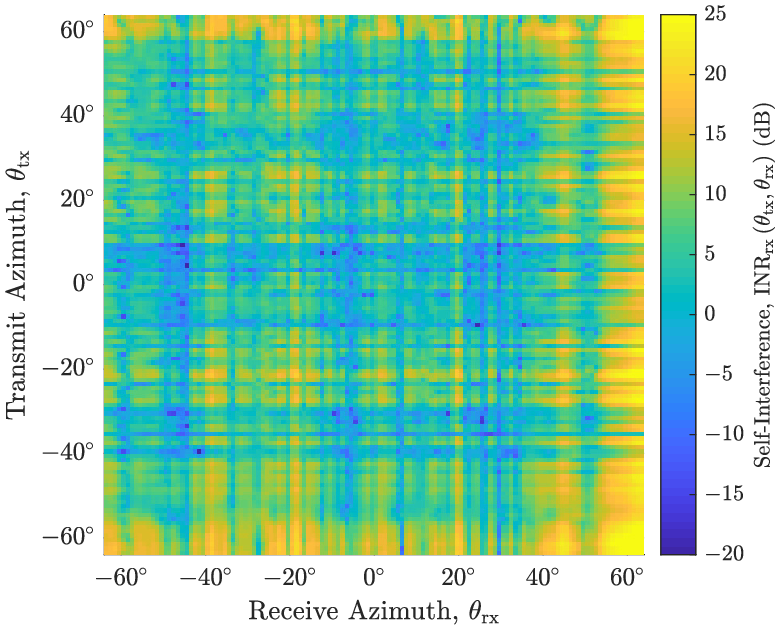}
        \label{fig:env-chamber}}
    \hfill
    \subfloat[Lab environment.]{\includegraphics[width=0.32\linewidth,height=\textheight,keepaspectratio]{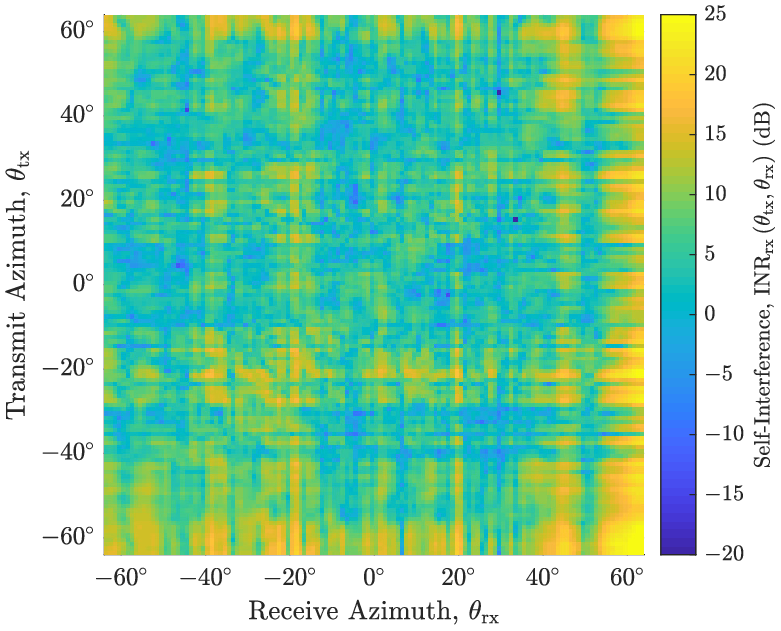}
        \label{fig:env-lab}}
    \hfill
    \subfloat[Lobby environment.]{\includegraphics[width=0.32\linewidth,height=\textheight,keepaspectratio]{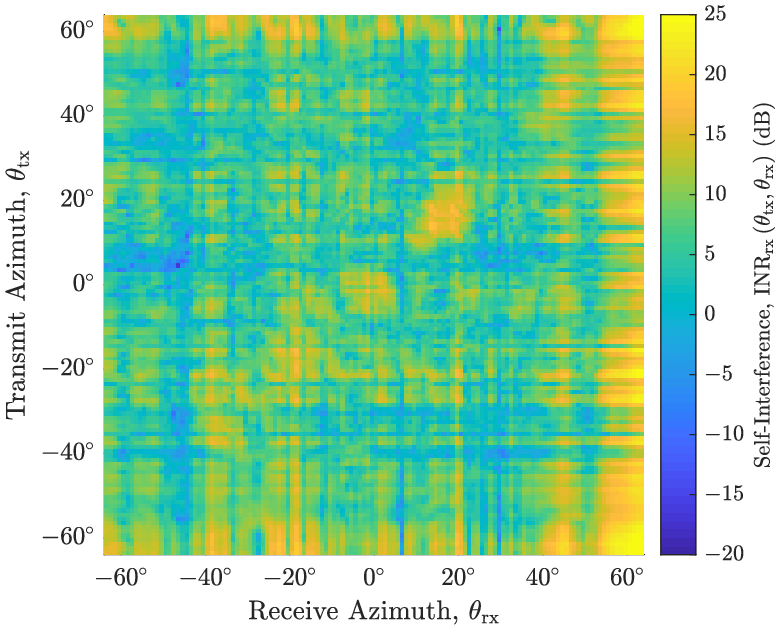}
        \label{fig:env-lobby}}
    \caption{Measured self-interference $\inrrx\pthtxrx$ as a function of transmit and receive steering direction $\pthtxrx$ in various environments. Direct coupling produces a stark grid-like structure that dominates, but reflections off the environment certainly contribute self-interference, especially along the diagonal where $\thtx \approx \thrx$.}
    \label{fig:env}
\end{figure*}

\begin{figure}
    \centering
    \includegraphics[width=0.4\linewidth,height=0.27\textheight,keepaspectratio]{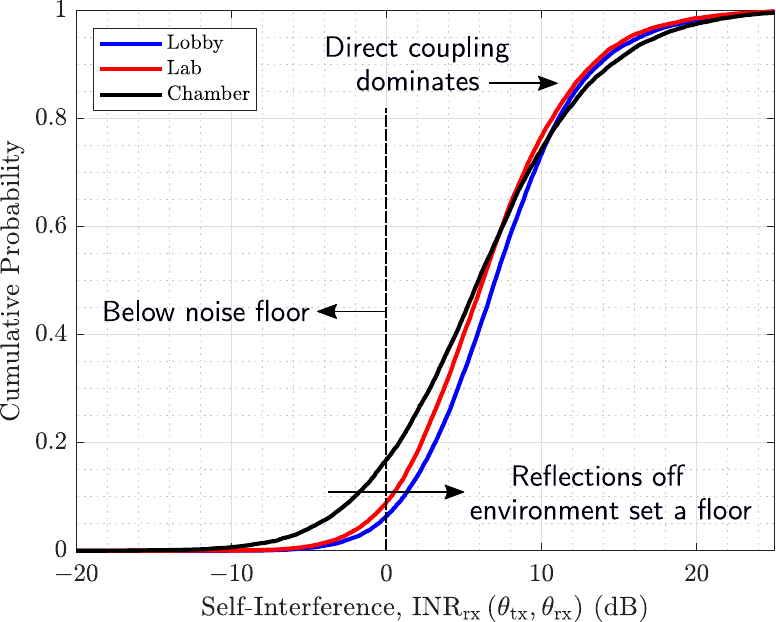}
    \caption{The \gpcdf of the measurements of self-interference shown in \figref{fig:env}. Reflections off the environment set a floor on the lower tail but do not impact the upper tail much, which largely stems from the direct coupling between the arrays.}
    \label{fig:env-cdf}
    \vspace{-0.25cm}
\end{figure}

In each of these three environments, we collected measurements of self-interference using the default configuration described prior, with each environment static during measurement.
The raw measurements of self-interference in each environment are shown in \figref{fig:env}, with their \gpcdf shown in \figref{fig:env-cdf}.
From the \gpcdf in \figref{fig:env-cdf}, over $80\%$ of measurements in the anechoic chamber yielded $\inrrx > 0$~dB, where self-interference is stronger than noise, with worst-case offenders exceeding well above $10$ dB toward $25$ dB.
Compared to the anechoic chamber, self-interference generally increased in the lab and lobby environments, visible in the rightward shift of the \gpcdf, with most of this being at the lower tail.

\begin{takeaway}
For most beam pairs, self-interference is too high\footnote{Typical full-duplex applications need roughly $\inrrx \leq 0$ dB to be worthwhile but exact requirements vary by application.} for worthwhile full-duplex operation, regardless of environment, motivating the need for additional cancellation of some form (e.g., beamforming, analog, and/or digital).
In \secref{sec:steer}, we introduce and experimentally evaluate a novel beamforming-based solution called \steerp to reduce self-interference.	
\end{takeaway}

From the measurements taken in the chamber, shown in \figref{fig:env-chamber}, we can see that there is strong self-interference along the bottom edge where $\thetatx \approx -60^\circ$ or the right edge where $\thetarx \approx 60^\circ$.
Conceptually, since the transmit array is to the right of the receive array (see \figref{fig:sweep}), this corresponds to when the transmit array steers its beam toward the receive array or when the receive array steers its beam toward the transmit array.  
As a result, self-interference is especially strong in the lower-right corner, where $\pthtxrx = (-60^\circ,60^\circ)$, since this corresponds to when the transmit array and receive array steer their beams toward one another.
The resulting grid-like structure is apparent, with low self-interference beam pairs sprinkled throughout, has noteworthy structural similarities to the measurements in \cite[Fig.~5a]{roberts_att_angular}, which were also taken in an anechoic chamber but with arrays oriented away from one another by $120^\circ$.
The lab and lobby settings, shown in \figref{fig:env-lab} and \figref{fig:env-lobby}, exhibit similar levels of self-interference as one another, and the grid-like structure is present in both but is less pronounced than in the chamber.


\begin{takeaway}
Reflections off the environment clearly contribute self-interference and are most recognizable along the diagonal when $\thtx \approx \thrx$, since this is where the transmit beam and receive beam are steered toward the same object in the environment.
Conceptually, reflections can contribute to off-diagonal entries (i.e., $\thetatx \not\approx \thetarx$) if they bounce off multiple objects or due to leakage in side lobes, but single-bounce reflections along $\thetatx \approx \thetarx$ appear to be more prominent---these undergo lower propagation/reflection losses and are illuminated and received by high-gain main lobes.
\end{takeaway}

\begin{takeaway}
    Beam pairs yielding low self-interference are less common in real-world settings, compared to in the anechoic chamber.
	In other words, reflections off the environment ``fill in the gaps'' to some degree, setting a floor of self-interference and pushing the lower tail of the \gcdf rightward.
	Nonetheless, the dominant source of self-interference is indeed the direct coupling between the arrays, visible in all three environments and in the upper tails of the \gpcdf, which is mainly when the transmit array steers its beam toward the receive array or when the receive array steers its beam toward the transmit array.
\end{takeaway}

\begin{figure*}
    \centering
    \subfloat[Separated by 10 cm.]{\includegraphics[width=0.32\linewidth,height=\textheight,keepaspectratio]{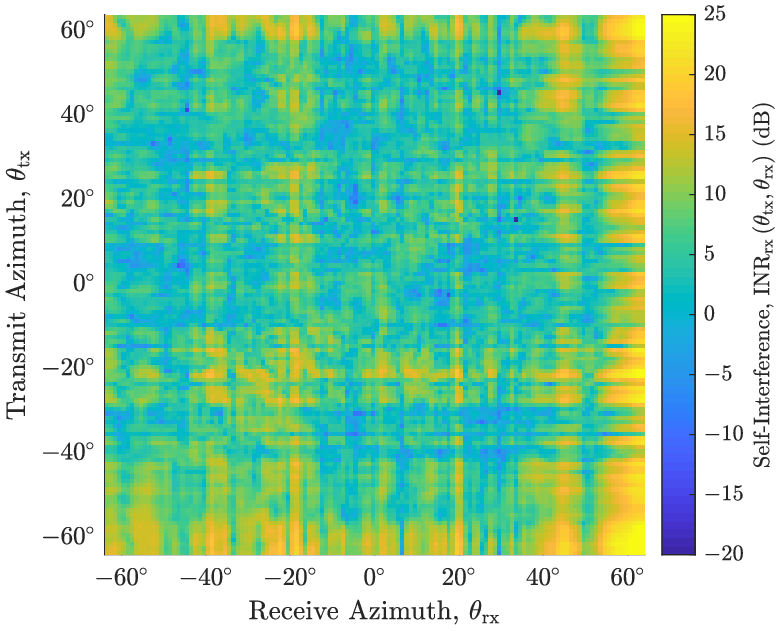}
        \label{fig:sep-a}}
    \hfill
    \subfloat[Separated by 20 cm.]{\includegraphics[width=0.32\linewidth,height=\textheight,keepaspectratio]{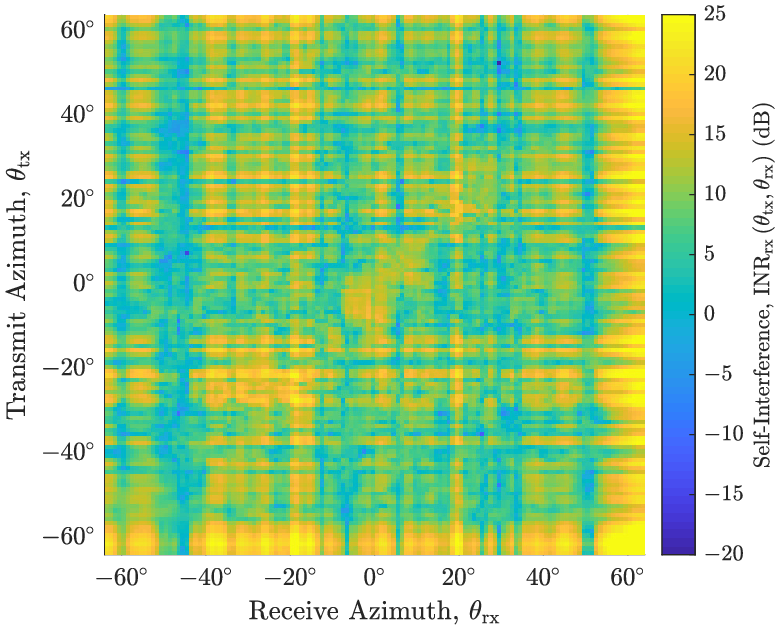}
        \label{fig:sep-b}}
    \hfill
    \subfloat[Separated by 30 cm.]{\includegraphics[width=0.32\linewidth,height=\textheight,keepaspectratio]{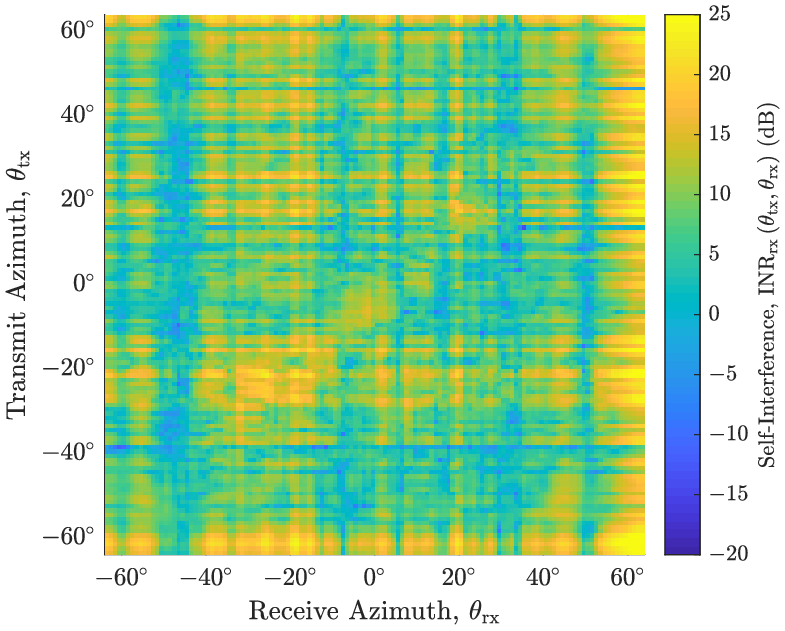}
        \label{fig:sep-c}}
    \caption{Measured self-interference $\inrrx\pthtxrx$ as a function of transmit and receive steering direction $\pthtxrx$ for various horizontal separations of the transmit and receive arrays. Measurements were taken in the lab environment.}    
    \label{fig:sep}
\end{figure*}

\subsection{Horizontal Separation of the Arrays}
\begin{figure*}
    \centering
    \subfloat[Horizontal separation.]{\includegraphics[width=0.4\linewidth,height=024\textheight,keepaspectratio]{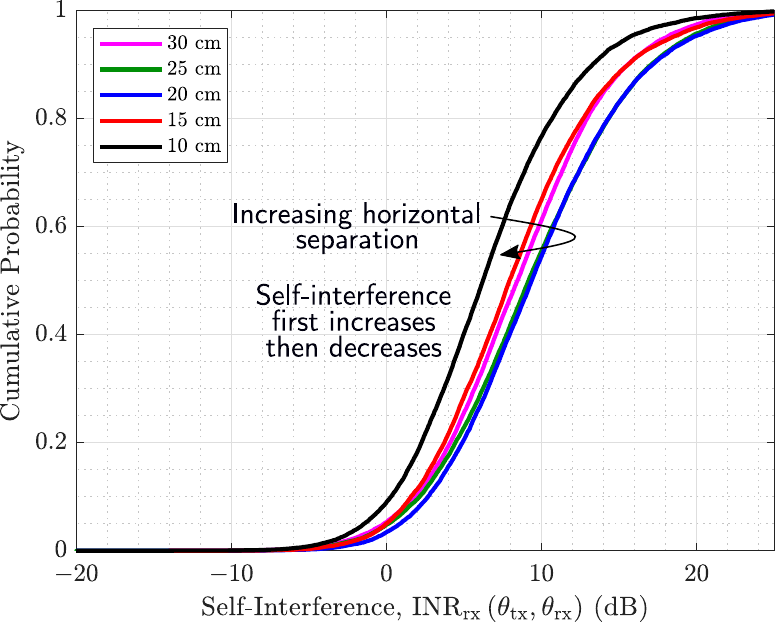}
        \label{fig:sep-cdf}}
    \qquad
    \subfloat[Angular separation.]{\includegraphics[width=0.4\linewidth,height=024\textheight,keepaspectratio]{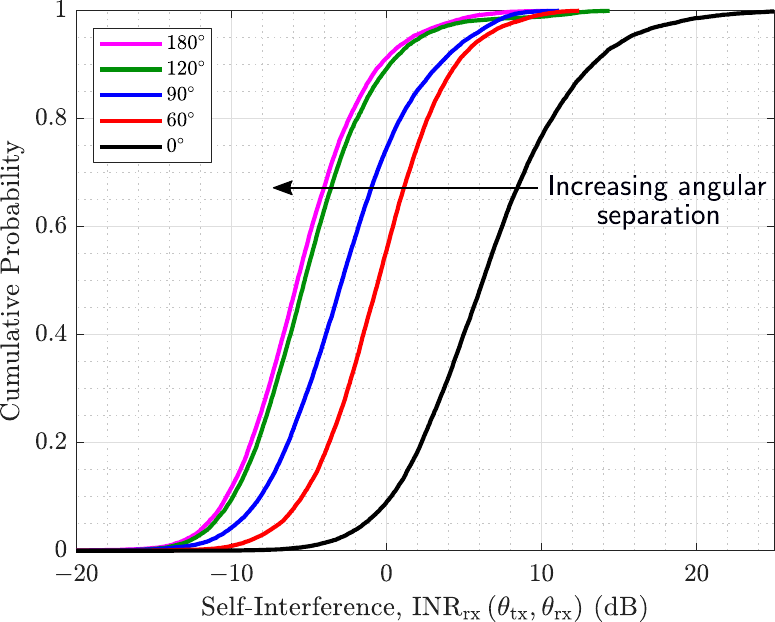}
        \label{fig:angle-cdf}}
    \caption{The \gpcdf of measured self-interference for various (a) horizontal separations and (b) angular separations of the transmit and receive arrays. Increasing horizontal separation does not necessarily result in decreased self-interference, whereas increasing angular separation reliably does.}
    \label{fig:sep-angle-cdf}
\end{figure*}

The measurements highlighted in the previous subsection were all taken with the transmit and receive arrays horizontally separated by 10 cm. 
Now, in the lab environment, we investigate the effects of increased horizontal separation on self-interference.
We do this by collecting measurements for horizontal separations of 10 cm, 15 cm, 20 cm, 25 cm, and 30 cm, whose \gpcdf are shown in \figref{fig:sep-cdf}. 
Notice that self-interference initially increases as separation is increased from 10 cm to 15 cm and on to 20 cm. 
It remains fairly unchanged as separation increases to 25 cm but then decreases when separation increases to 30 cm.
This is further explored in \figref{fig:sep}, which shows self-interference for separations of 10 cm, 20 cm, and 30 cm.
The spatial structure appears to be fairly similar across all three, but certain spatial components become more prominent as separation is increased.



\begin{takeaway}
This non-monotonic nature as a function of horizontal separation perhaps best highlights the non-trivial spatial composition of the self-interference channel $\mH$.
In other words, increasing the arrays' separation does not merely increase over-the-air path loss but rather changes the channel's composition to some degree, though it is difficult to concretely explain how so.
\end{takeaway}


\subsection{Angular Separation of the Arrays}
Thus far, we have only considered transmit and receive arrays oriented in the same plane, without any angular separation.
We now explore how the self-interference profile varies with increased angular separation of the arrays.
In the lab environment, we collected measurements when the transmit and receive arrays were separated by $0^\circ$, $60^\circ$, $90^\circ$, $120^\circ$, and $180^\circ$.
In all cases, the centers of the arrays were approximately 10~cm apart.
An angular separation of $120^\circ$, for example, may correspond to outward-facing arrays mounted on separate sides of a triangular-shaped, sectorized \bs. 
The \gcdf of self-interference for each orientation is shown in \figref{fig:angle-cdf}.
There is a noteworthy decrease in self-interference of about $7$ dB in median when going from $0^\circ$ to $60^\circ$. 
By $90^\circ$, around $75\%$ of self-interference measurements are below the noise floor.
This trend continues with $120^\circ$ and $180^\circ$, which produce distributions of self-interference comparable to one another that lay above $0$ dB for around only $10\%$ of measurements.

\begin{takeaway}
    Clearly, angular separation can be a powerful means to reduce self-interference, reducing it to below the noise floor in our measurements, and these results may be particularly relevant to sectorized \bss where \mmwave transceivers are mounted on separate sides of a triangular or square platform.
	It is important to note that, increasing angular separation may not reduce self-interference to below the noise floor in other systems, especially those transmitting with higher power. 
    For instance, even with $120^\circ$ separation, the vast majority of measurements in \cite{roberts_att_angular} saw self-interference well above the noise floor.
\end{takeaway}

\begin{figure*}
    \centering
    \subfloat[Separated by $60^\circ$.]{\includegraphics[width=0.32\linewidth,height=\textheight,keepaspectratio]{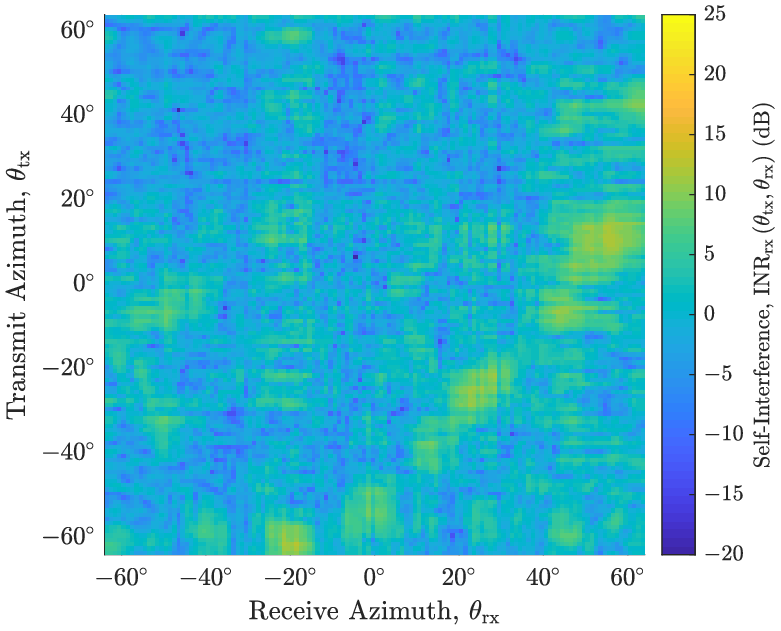}
        \label{fig:angle-60}}
    \hfill
    \subfloat[Separated by $90^\circ$.]{\includegraphics[width=0.32\linewidth,height=\textheight,keepaspectratio]{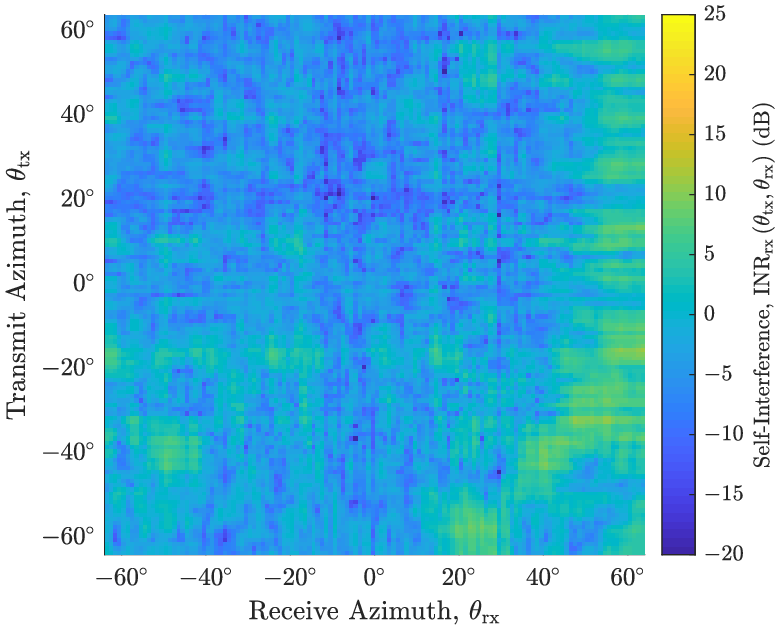}
        \label{fig:angle-90}}
    \hfill
    \subfloat[Separated by $120^\circ$.]{\includegraphics[width=0.32\linewidth,height=\textheight,keepaspectratio]{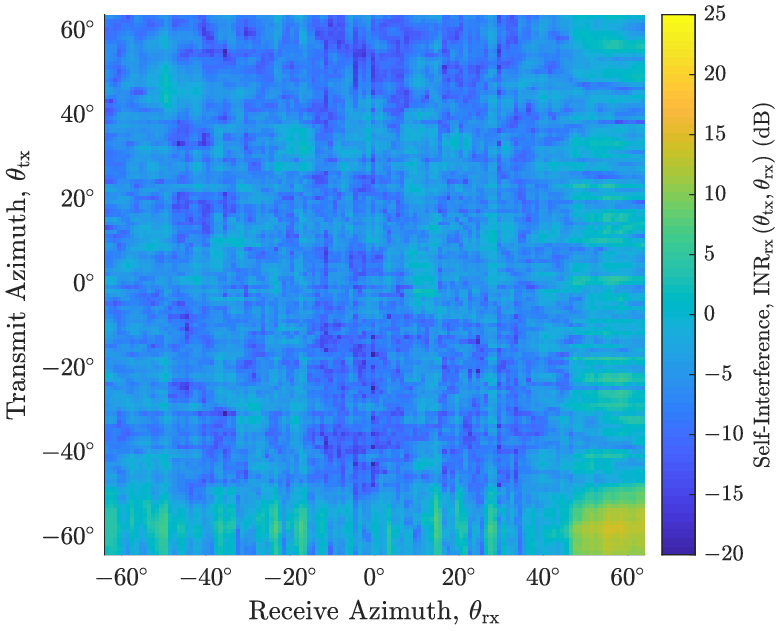}
        \label{fig:angle-120}}
    \caption{Measured self-interference for various angular separations of the transmit and receive arrays in the lab environment. The presence of direct coupling has weakened, and reflections off the environment manifest along diagonals shifted downward and rightward in accordance with the angular separation.}    
    \label{fig:angle}
\end{figure*}


Now, consider \figref{fig:angle}, which depicts measured self-interference as a function of transmit and receive direction for angular separations of $60^\circ$, $90^\circ$, and $120^\circ$.
For comparison purposes, we have kept the color scale the same as all previous plots, which immediately conveys the stark decrease in self-interference.
The direct coupling between the arrays has decreased significantly, presumably in part due to the isolation provided by the metal plates on which our arrays are mounted, as visible in \figref{fig:arrays}.

\begin{takeaway}
	One observation to make is that the strong self-interference due to reflections---previously along the diagonal $\thetatx \approx \thetarx$---shifts downward and rightward with increased angular separation. 
	This is because the transmit and receive beams now steer toward the same reflective object in the environment when their steering directions are offset by this angular difference. 
	For instance, with $90^\circ$ of separation, the beams steer toward the same object when $\thetarx \approx \thetatx + 90^\circ$.
\end{takeaway}

\comment{
\subsection{Metal Isolator Between the Arrays}
As opposed to mere \textit{spatial} isolation, we now investigate the effect a metal isolator placed between the arrays has on self-interference.
To do this, we separated the transmit and receive arrays by 20 cm in the lab environment and took measurements with and without a metal plate between the arrays.
The metal plate used was a spare backplate used to mount the 60 GHz arrays, as visible in \figref{fig:arrays}.
The field-of-view of each array was not impaired, as the plate was positioned to not extend past the actual array elements.
A comparison of self-interference with and without the metal isolator is shown in \figref{fig:iso-diff} while their \gpcdf are shown in \figref{fig:iso-cdf}.

\begin{figure*}
    \centering
    \subfloat[Comparing distributions.]{\includegraphics[width=\linewidth,height=0.22\textheight,keepaspectratio]{plots/main_wilab_v03_07}
        \label{fig:iso-cdf}}
    \qquad
    \subfloat[Gain in self-interference with isolator.]{\includegraphics[width=\linewidth,height=0.22\textheight,keepaspectratio]{plots/main_wilab_v03_08-3}
        \label{fig:iso-diff}}
    \caption{(a) The \gcdf of measured self-interference with and without a metal isolator between the transmit and receive arrays in the lab environment. (b) The gain in self-interference with the metal isolator, relative to self-interference without the isolator.} 
    \label{fig:iso}
\end{figure*}

From \figref{fig:iso-cdf}, we can see that the metal isolator offers a reduction of about $3$ dB in median self-interference.  
To more closely examine this, in \figref{fig:iso-diff}, we plot the gain in self-interference with the isolator, relative to self-interference without the isolator.
The spatial components along the top, right, and bottom edges---which we attributed to the direct coupling based on \figref{fig:env-chamber}---all see interference reduced by about $4$--$6$ dB.
The spatial components along the diagonal---from reflections in the environment---see little to no reduction in self-interference, as intuition suggests.
Interestingly, a few spatial components see minor increases in self-interference with the isolator, likely due to slight, non-trivial changes in the channel composition that results in slightly more constructive coupling of the beams.
}

\begin{figure*}
    \centering
    \subfloat[Transmit on right, receive on left.]{\includegraphics[width=0.34\linewidth,height=0.27\textheight,keepaspectratio]{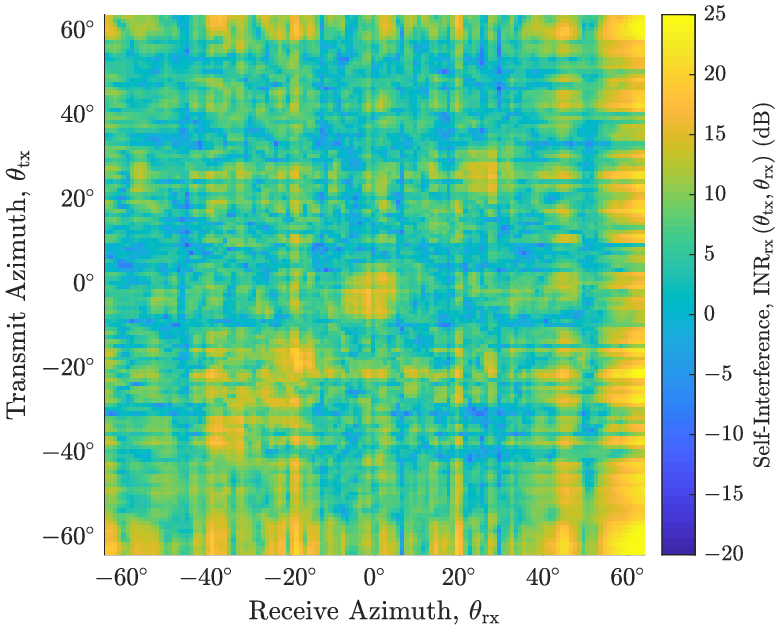}
        \label{fig:recip-a}}
    \qquad
    \subfloat[Transmit on left, receive on right.]{\includegraphics[width=0.34\linewidth,height=0.27\textheight,keepaspectratio]{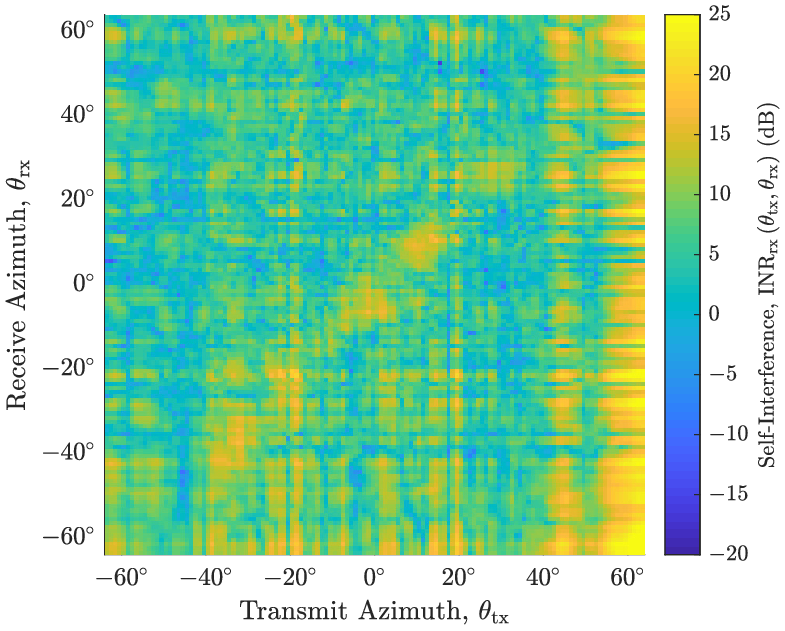}
        \label{fig:recip-b}}
    \caption{Comparing the spatial profile of self-interference when the transmit and receive array positions are swapped. Both were taken in the lab environment with the arrays separated by 10 cm and with no angular separation. Note that the axes of (b) have been swapped for visualization.}
    \label{fig:recip}
\end{figure*}

\subsection{Confirming Transmit-Receive Reciprocity}
Until now, all measurements have been taken with the transmit array on the right and receive array on the left, as shown in \figref{fig:sweep}.
We now swap the transmit and receive arrays and repeat our measurements to compare self-interference profiles.
In both cases, the transmit and receive arrays were separated by 10 cm with no angular separation and placed in the lab environment.
In \figref{fig:recip}, we plot our measurements of self-interference for both orientations; notice that \figref{fig:recip-b} has its axes swapped to more easily compare it to \figref{fig:recip-a}.
It is immediately clear that a strong sense of reciprocity exists, considering the stark similarity between the two sets of measurements.
The direct coupling along the right and bottom edges appears prominently in both, with even some of the very fine artifacts present in both.
The most significant difference between the two is perhaps along the diagonal $\thtx \approx \thrx$.
While quite similar, there are some differences, which are most likely attributed to the fact that the transmit and receive arrays' perspective of the environment slightly changes when their positions are swapped; 
this may lead to signals not illuminating or reflecting off objects \textit{exactly} the same in both orientations.

\begin{takeaway}
Overall, this strong reciprocity is somewhat expected but is nonetheless a welcome sight.
In cases where the arrays' roles may not be dedicated to transmission or reception, such as in a sectorized \bs, this reciprocity of self-interference may be exploited to reduce the configuration overhead of full-duplex solutions since measurements may only need to be taken in one direction.
\end{takeaway}


\subsection{Small-Scale Spatial Variability} \label{subsec:small-scale}
Measurements in prior work \cite{roberts_att_angular} showed that slightly shifting the steering directions of the transmit and receive beams can drastically alter the degree of self-interference coupled.
This was an important finding that was the motivation behind follow-on work \cite{roberts_steer}. 
With this in mind, we have conducted a similar analysis as in \cite{roberts_att_angular} to investigate the presence of this small-scale spatial variability of self-interference in our system.

To do this, we processed our measurements similar to what was done in \cite{roberts_att_angular}.
We began by forming the so-called transmit and receive \textit{spatial neighborhoods} surrounding the $i$-th transmit direction $\thtxi \in \txdirsetcb$ and the $j$-th receive direction $\thrxj \in \rxdirsetcb$.
The transmit spatial neighborhood $\txdirsetmeas\idx{i}$ is defined as the set of all measured directions within some azimuthal distance $\Delta\thetatx$ of $\thtxi$.
Defining the receive spatial neighborhood analogously, we can write these neighborhoods as
\begin{align}
\txdirsetmeas\idx{i}\parens{\Delta\thetatx} &= \braces{\thetatx \in \txdirsetcb : \anglediff{\thetatx,\thetatx\idx{i}} \leq \Delta\thetatx} \\
\rxdirsetmeas\idx{j}\parens{\Delta\thetarx} &= \braces{\thetarx \in \rxdirsetcb : \anglediff{\thetarx,\thetarx\idx{j}} \leq \Delta\thetarx},
\end{align}
where $\anglediff{\theta_1,\theta_2}$ is the absolute difference between two angles $\theta_1$ and $\theta_2$.
Then, the set of all self-interference measurements over these spatial neighborhoods can be formed as
\begin{align}
\Iij\parens{\Delta\thetatx,\Delta\thetarx} = \braces{\inrrx\pthtxrx : \thtx \in \txdirsetmeas\idx{i}\parens{\Delta\thetatx}, \thrx \in \rxdirsetmeas\idx{j}\parens{\Delta\thetarx}}.
\end{align}
The minimum self-interference observed over these spatial neighborhoods is
\begin{align}
\inrijmin\parens{\Delta\thetatx,\Delta\thetarx} &= \minop{\Iij\parens{\Delta\thetatx,\Delta\thetarx}},
\end{align}
while the range in self-interference over the neighborhood we define as
\begin{align}
\inrijrng\parens{\Delta\thetatx,\Delta\thetarx} &= \frac{\maxop{\Iij\parens{\Delta\thetatx,\Delta\thetarx}}}{\minop{\Iij\parens{\Delta\thetatx,\Delta\thetarx}}}.
\end{align}
Here, $\inrijmin\parens{\Delta\thetatx,\Delta\thetarx}$ is the minimum self-interference that can be reached if the $i$-th transmit beam and $j$-th receive beam are allowed to shift their steering directions by at most $\Delta\thetatx$ and $\Delta\thetarx$, respectively.
The range $\inrijrng\parens{\Delta\thetatx,\Delta\thetarx}$ captures how much absolute spatial variability in self-interference exists over the neighborhood.

\begin{figure*}
    \centering
    \subfloat[Range in self-interference, $\inrijrng$.]{\includegraphics[width=0.4\linewidth,height=\textheight,keepaspectratio]{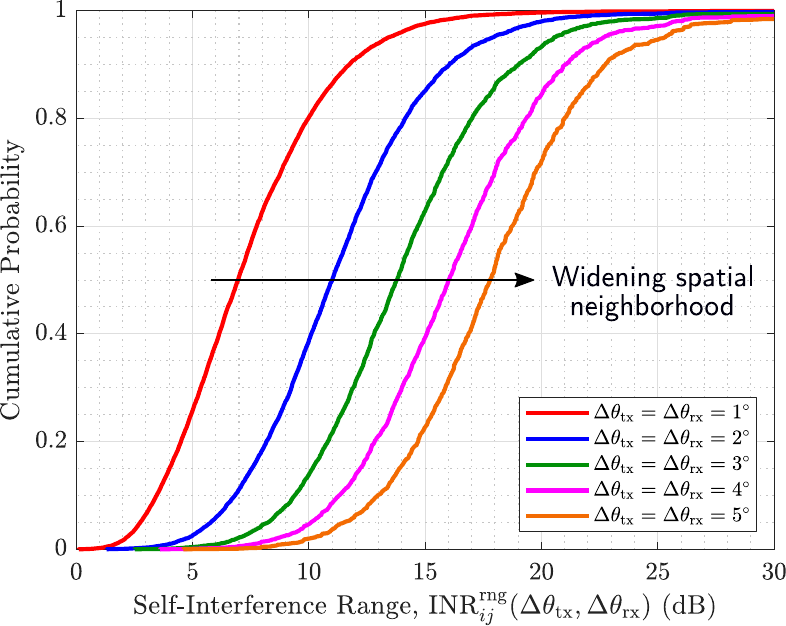}
        \label{fig:inr-rng}}
    \qquad
    \subfloat[Minimum self-interference, $\inrijmin$.]{\includegraphics[width=0.4\linewidth,height=\textheight,keepaspectratio]{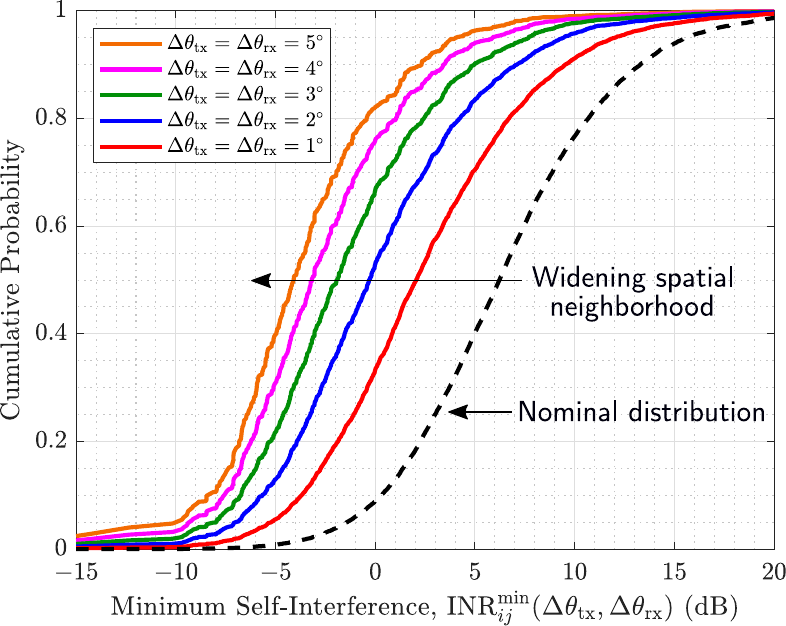}
        \label{fig:inr-min}}
    \caption{The \gpcdf of (a) self-interference range $\inrijrng\parens{\Delta\thetatx,\Delta\thetarx}$ and (b) minimum self-interference $\inrijmin\parens{\Delta\thetatx,\Delta\thetarx}$ for various neighborhood sizes $\parens{\Delta\thetatx,\Delta\thetarx}$. The \gpcdf are taken across all measured transmit and receive beam pairs $\parens{\thtxi,\thrxj}$.} 
    \vspace{-0.65cm}
    \label{fig:inr-rng-min}
\end{figure*}

In \figref{fig:inr-rng}, we plot the \gcdf of $\inrijrng$ for various neighborhood sizes $\parens{\Delta\thetatx,\Delta\thetarx}$, where each \gcdf is over all measured transmit-receive steering directions $\parens{\thtxi,\thrxj}$.
Likewise, we plot the \gpcdf of $\inrijmin$ in \figref{fig:inr-min}.
These are analogous to those produced in \cite[Figs.~11a, 13a]{roberts_att_angular}.
From \figref{fig:inr-rng}, we can see that, for $50$\% of transmit-receive beam pairs, self-interference can vary by over $11$ dB if the beams shift by at most $2^\circ$.
A small fraction of beam pairs are quite sensitive, varying by more than $15$ dB if the beams shift by at most $1^\circ$.
On the other hand, the lower tail represents beam pairs that are less sensitive, surrounded by beam pairs yielding similar levels of self-interference.

In \figref{fig:inr-min}, we gain insight into the potential to reduce self-interference through small shifts of the transmit and receive beams.
Without shifting the beams, the nominal distribution of self-interference is shown as the dashed line, laying above $0$ dB over $90$\% of the time; this is the same as the lab's distribution shown in \figref{fig:env-cdf}.
If permitting the transmit and receive beams to shift, low self-interference can be reached much more often.
With at most $1^\circ$ of shifting, over $30$\% of beam pairs can reach $\inrrx < 0$ dB.
With at most $3^\circ$, two-thirds of beam pairs can reach $0$~dB or less.
Note that we observe less small-scale spatial variability in our measurements than that observed in \cite[Figs.~11a, 13a]{roberts_att_angular}, which is presumably attributed to the fact that our beams were swept only in azimuth using linear arrays whereas both azimuth and elevation were swept in \cite{roberts_att_angular} using planar arrays, leading to greater spatial richness.

\begin{takeaway}
Slight shifts of the transmit and receive steering directions (on the order of one degree) can lead to significant fluctuations in self-interference. 
These shifts are small relative to the size of the main lobe (our half-power beamwidth is around $10^\circ$), which suggests that this small-scale spatial variability may be exploited to create full-duplex solutions, such as \steer \cite{roberts_steer}, that reduce self-interference via slight shifts of the transmit and receive beams while preserving high \gpsnr on the downlink and uplink.
We explore this more in \secref{sec:steer}.
\end{takeaway}



\section{Key Challenges and a Framework for Practical Beamforming Solutions} \label{sec:challenges}

From the measurements presented in the previous section, we observed that self-interference is typically prohibitively high in full-duplex \mmwave systems in a variety of environments and configurations, even with directional beamforming.
This motivates the need to take additional measures to mitigate self-interference to enable full-duplex \mmwave systems.
As mentioned in the introduction, a considerable amount of recent work has proposed designing transmit and receive beams at a \mmwave transceiver in such a way that cancels self-interference spatially to enable full-duplex operation.
Most of these existing solutions, however, would face significant hurdles if directly translated to real-world systems.
In this section, we outline key challenges that practical full-duplex \mmwave systems would face, in an effort to steer and motivate future research.
We then conclude this section by proposing a general framework that can guide the development of practical beamforming-based solutions for full-duplex \mmwave systems.


\begin{challenge}[Downlink and Uplink Channel Knowledge is Unlikely]
A number of existing beamforming solutions for full-duplex rely on estimation of the downlink and uplink \mimo channels between a full-duplex \bs and the users it aims to serve (i.e., $\vhtx$, $\vhrx$).
In general, such channel estimation is unlikely in practical systems, largely due to the sheer size of \mmwave \mimo channels and the fact that these channels are typically observed through the lens of analog beamforming.
As such, practically-sound full-duplex solutions for today's \mmwave systems should ideally not rely on full knowledge of downlink/uplink \mimo channels.
\end{challenge}

\begin{challenge}[Accommodating Beam Alignment]
Instead of estimating high-dimensional downlink/uplink \mimo channels, practical \mmwave systems (e.g., 5G, IEEE 802.11ay) have turned to \textit{beam alignment} to determine which beam a \bs uses to serve each \ue \cite{ethan_beam}.
Through a series of \gls{rsrp} measurements taken with candidate beams from some codebook, a \mmwave \bs can identify beams that deliver high beamforming gain to each user.
Albeit simplistic, codebook-based beam alignment has proven to be a robust mechanism to close the link between a \bs and \ue while also facilitating initial access.
Practical beamforming-based solutions for full-duplex \mmwave systems should therefore strive to accommodate beam alignment to ensure they more seamlessly integrate into practical systems.
\end{challenge}

\begin{challenge}[Self-Interference Channel Estimation is Difficult]
Like that of the downlink and uplink channels, the potential for self-interference \mimo channel estimation is also unlikely in full-duplex \mmwave systems.
Again, this is largely due to the sheer size of the channel and complications imposed by analog beamforming.
We believe solutions that instead rely on measurements of self-interference for \textit{particular} transmit and receive beams---as opposed to estimation of the entire \mimo channel---will be more practically viable and robust.
This is reinforced by the fact that minor self-interference channel estimation errors could be detrimental to full-duplex performance.
Nonetheless, accurate modeling and/or thorough estimation of the direct coupling and static environment (e.g., buildings, infrastructure) could certainly be useful. 
\end{challenge}

\begin{challenge}[Time-Variability of Self-Interference]
Naturally, as the environment changes, the self-interference profile will vary to some degree and full-duplex solutions will need to be adaptive to ensure they remain effective. 
At the same time, full-duplex solutions should strive to minimize their consumption of radio resources during this adaptation. 
As a result, if self-interference varies rapidly on the time-scale of data transmissions, it may be challenging to create reliable, low-overhead full-duplex solutions.
The time-variability of self-interference in \mmwave systems has not been extensively investigated, making it a good topic for future work.
\end{challenge}

\begin{challenge}[Beamforming in Practical Phased Arrays]
    Even if one were to design an ideal beamforming-based solution in theory based on perfect channel knowledge, realizing the design in practice may not be as straightforward as one may expect for a few reasons.
    Practical \mmwave transceivers often rely on analog beamforming as a cost- and power-efficient architecture to electronically steer beams with dense antenna arrays.
    While hybrid digital/analog beamforming is also an attractive architecture conceptually, it is more difficult to implement in practice, with fully-connected architectures being especially challenging.
    Given this, full-duplex solutions that are suitable for \textit{analog-only} beamforming systems will be more practically relevant, and in fact, such solutions could also be directly applied to some hybrid beamforming systems.
    In such analog beamforming architectures, \textit{digitally-controlled} \rf components such as phase shifters, attenuators, and amplifiers are used to shape beams.
    As a result, the control resolution of each is limited, and there is typically some error margin and frequency-selectivity associated with each component.
    Electromagnetic coupling, imperfect phase calibration across elements, and non-isotropic antenna elements can all lead to additional challenges when realizing a desired beamforming design in actual hardware.
    Combining all of this with the fact that small errors in cancelling self-interference can be detrimental to full-duplex system performance complicates the real-world implementation of sophisticated beamforming-based solutions. 
    This makes it attractive to explicitly \textit{measure} the performance of candidate beams to verify they couple low self-interference and/or deliver high beamforming gain on the downlink and uplink before relying on them for data transmission.
    Perhaps machine learning is one route to overcoming some of these challenges of real-world phased arrays, as demonstrated in \cite{yu_nulling_2022,asu_adapt_2022,yu_rl_2022}. 
\end{challenge}

\subsection{Proposed Framework}

\begin{figure*}
    \centering
    \includegraphics[width=\linewidth,height=0.18\textheight,keepaspectratio]{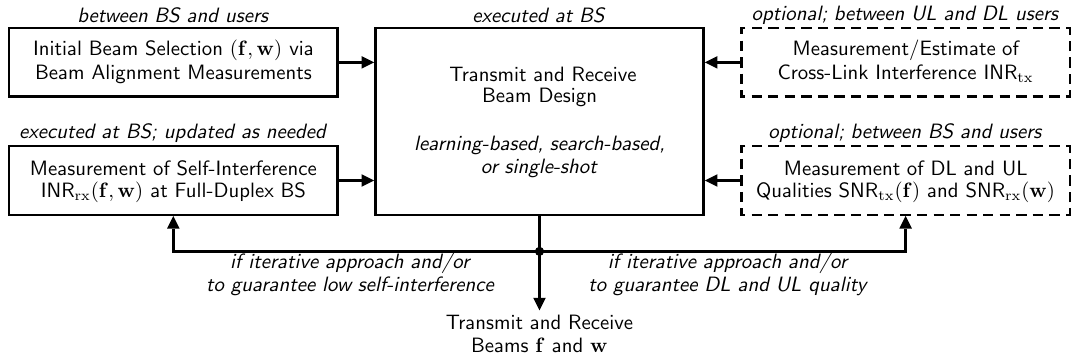}
    \caption{A block diagram of our proposed measurement-driven framework for designing beamforming-based solutions for full-duplex \mmwave systems. Initialized by beam selections following beam alignment, transmit and receive beams can be designed based on measurements of self-interference, downlink \gsnr, and uplink \gsnr, circumventing the need for \mimo channel estimation. Both \steer \cite{roberts_steer} and \steerp (presented in \secref{sec:steer}) comply with this framework.}
    \label{fig:framework}
\end{figure*}

In order to address some of the aforementioned challenges, we now propose a general framework for beamforming-based full-duplex solutions.
Our framework, illustrated in \figref{fig:framework}, consists of five high-level components.
As mentioned previously, current \mmwave systems rely on beam alignment to close the link between devices.
To accommodate this flexibly, we suggest that the design of beamforming-based solutions for full-duplex be initialized by beam selection following beam alignment measurements, as shown in the upper left box of \figref{fig:framework}.
This will help maintain backwards-compatibility with existing standards such as 5G and IEEE 802.11ay.

Shown as the central box, the design of transmit and receive beams at the full-duplex \bs will take these initial beam selections as input along with three other components.
Perhaps this design is powered by a learning framework (e.g., \cite{yu_nulling_2022,yu_rl_2022}) and/or is conducted in an iterative fashion (e.g., \cite{roberts_steer}), where it sequentially produces beam candidates.
As shown in the lower left box, we propose that the design be driven by measurements of self-interference---specifically $\inrrx\parens{\vf,\vw}$ for some transmit and receive beams---as opposed to estimation of the self-interference channel $\mH$.
Note that these measurements may not all be taken collectively at once but rather as necessary one-by-one.
Moreover, these measurements only need to be updated as necessary, depending on the time-variability of self-interference, and can be referenced otherwise.
Optionally, a beamforming design also may take into account either measurement or an estimate of cross-link interference $\inrtx$ between the uplink and downlink users.
If measured, this must be fed back to the \bs; an estimate is therefore likely preferable in practice.

Suppose the \bs designs some transmit and receive beams $(\vf,\vw)$ that it intends to use for full-duplexing downlink and uplink. 
To ensure that these beams deliver sufficient quality, it may be worthwhile to explicitly measure $\snrtx\parens{\vf}$ and $\snrrx\parens{\vw}$ with these beams.
If the measured \gpsnr are satisfactory, perhaps they can then be used for service.
If not, the \bs may need to design new beam candidates $\parens{\vf,\vw}$ that can then be verified as delivering sufficiently high $\snrtx\parens{\vf}$ and $\snrrx\parens{\vw}$. 
Note that this process can also be used to ensure that self-interference $\inrrx\parens{\vf,\vw}$ is sufficiently low for particular beam candidates.

Ideally, beam design should be executed at the full-duplex \bs and should aim to minimize the radio resources it demands.
Measurement between the \bs and \ues should be minimized, along with any over-the-air feedback.
Note that measurements of self-interference take place entirely at the \bs, meaning they introduce no feedback.
Nonetheless, the number of self-interference measurements should be minimized.
Perhaps an iterative approach would be ideal to avoid taking more measurements than necessary, and attractive solutions would leverage any long-term statistics of self-interference to accelerate design and reduce measurement overhead.






\section{Real-World Evaluation of Beamforming-Based Full-Duplex Solutions} \label{sec:steer}
In this section, we perform a real-world evaluation of two beamforming-based solutions for full-duplex \mmwave systems.
Using our 60 GHz phased arrays, we first implement prior work \steer~\cite{roberts_steer} to evaluate its effectiveness as a full-duplex solution in a real-world setting.
Then, based on the framework proposed in the previous section, we introduce a more robust version of \steer, which we call \steerp, to illustrate how the optional measurements of $\snrtx$ and $\snrrx$---the bottom right box of \figref{fig:framework}---can be used to improve system performance by better guaranteeing downlink and uplink quality in pursuit of reduced self-interference.

\subsection{An Existing Solution, \steer}
Recent work \cite{roberts_steer} proposes \steer, a beam selection methodology that reduces self-interference while delivering high beamforming gain on the downlink and uplink.
\steer achieves this by first conducting beam alignment to identify initial steering directions and then slightly shifting these steering directions (on the order of one degree) in search of lower self-interference.
Inspired by the small-scale spatial variability in \cite[Fig.~13a]{roberts_att_angular}---and seen in \figref{fig:inr-min}---\steer's working principle is that slight shifts of the beams will preserve high \gsnr on the downlink and uplink and significantly reduce self-interference, yielding \gpsinr sufficiently high for full-duplex. 

In \cite{roberts_steer}, the authors evaluated \steer using actual measurements of self-interference, but these were collected in an anechoic chamber, not in a real-world environment.
Moreover, the evaluation in \cite{roberts_steer} uses simulation to validate downlink and uplink quality, meaning their evaluation may not have accurately captured the true \gsnr loss when shifting the transmit and receive beams---an artifact of both real-world channels and real-world phased arrays.
In addition, \steer was only evaluated using a single 28 GHz phased array platform in a fixed orientation, meaning it is unclear if its success will generalize to other systems.
Altogether, this has motivated us to implement and experimentally evaluate the performance of \steer using our off-the-shelf phased arrays in a real-world environment with actual downlink and uplink users.
To our knowledge, this is the first experimental evaluation of a beamforming-based solution for full-duplex \mmwave systems, complete with self-interference, downlink, uplink, and cross-link interference measurements.

When serving a downlink user and an uplink user, \steer begins by using beam alignment measurements to solve (or approximately solve) the \gsnr-maximizations
\begin{align}
\thtxinit &= \argmax_{\thetatx \in \sCtx} \ \snrtx\parens{\thetatx}, \qquad
\thrxinit = \argmax_{\thetarx \in \sCrx} \ \snrrx\parens{\thetarx}, 
\end{align}
where $\sCtx$ and $\sCrx$ are beam codebooks used for beam alignment.
Note that these codebooks and the methods for conducting beam alignment are not within the scope of \steer since it can be applied atop any initial beam selections.
Conducting beam alignment yields directions in which the \bs can steer its beams to deliver high $\snrtx$ and $\snrrx$; these are used to initialize \steer.
Then, surrounding these initial steering directions, \textit{spatial neighborhoods} are populated as\footnote{Note that our phased arrays are \ulas, meaning we only consider the azimuth dimension, whereas  \cite{roberts_steer} employs planar arrays and considers both azimuth and elevation. As a result, there are minor differences in our presentation of \steer herein.}
\begin{align}
\thtxinit + \setnbr\parens{\Delta\thetatx,\delta\thetatx}, \qquad
\thrxinit + \setnbr\parens{\Delta\thetarx,\delta\thetarx}.
\end{align}
Here, the spatial neighborhood of size $\Delta\theta$ and resolution $\delta\theta$ is defined as
\begin{align}
\setnbr\parens{\Delta\theta,\delta\theta} 
&= \braces{m \cdot \delta\theta : m \in \brackets{-\floor{\frac{\Delta\theta}{\delta\theta}},\floor{\frac{\Delta\theta}{\delta\theta}}}},
\end{align}
where $\floor{\cdot}$ is the floor operation and $[a,b] \triangleq \braces{a,a+1,\dots,b-1,b}$.
An illustration of this concept of a spatial neighborhood can be found in \cite[Fig.~3]{roberts_steer}.

To refine its steering directions from the initial ones, the full-duplex \bs executes \steer by solving the following problem.
\begin{subequations}\label{eq:steer}
    \begin{align}
    \min_{\thtx,\thrx} \ \min_{\Delta\varthetatx,\Delta\varthetarx} \ & \Delta\varthetatx^2 + \Delta\varthetarx^2 \\ 
    \st \ 
    & \inrrx\thtxrx \leq \maxop{\inrrxtgt,\inrrxmin} \label{eq:steer-target} \\
    &\thtx \in \thtxinit+ \setnbr\parens{\Delta\varthetatx,\delta\thetatx} \\
    &\thrx \in \thrxinit + \setnbr\parens{\Delta\varthetarx,\delta\thetarx} \\
    & 0 \leq \Delta\varthetatx \leq \Delta\thetatx, \ 0 \leq \Delta\varthetarx \leq \Delta\thetarx
    \end{align}
\end{subequations}
In this problem, \steer aims to find the transmit and receive steering directions $\pthtxrx$ that yield self-interference $\inrrx\pthtxrx$ below some target $\inrrxtgt$; note that the $\max$ operation in \eqref{eq:steer-target} is simply to ensure the problem is feasible, where $\inrrxmin$ is the minimum self-interference over the neighborhood.
In an effort to preserve high \gpsnr, \steer minimizes the deviation of the transmit and receive beams from the initial selections $\parens{\thtxinit,\thrxinit}$ by minimizing the size of the spatial neighborhoods from which beams are selected, upper-bounded by $\Delta\thetatx$ and $\Delta\thetarx$.

Conceptually, \steer's beam-shifting approach seems sensible based on the small-scale spatial variability observed herein (\figref{fig:inr-min}) and in \cite{roberts_att_angular}.
However, such an approach may not always be reliable in real-world settings. 
First of all, if the initial beam selections are not well-aligned with the users, slights shifts could lead to prohibitive beamforming losses. 
In addition, shifting a beam's steering direction on the order of one degree is not necessarily straightforward in practical phased arrays; 
this can lead to minor losses in main lobe gain and hence losses in downlink/uplink \gsnr.
Additionally, if there is not rich spatial variability, \steer may require significant shifting to reduce self-interference, which can degrade downlink and uplink \gpsnr. 
As we will see, \steer can also be sensitive to neighborhood size $\parens{\Delta\thetatx,\Delta\thetarx}$, a design parameter; a neighborhood that is too small can restrict \steer from finding low self-interference, whereas one too large may lead to too much \gsnr degradation.
The optimal neighborhood size can vary for different downlink-uplink user pairs, as we will see shortly, making it unclear how to choose a neighborhood size that generalizes well.
Nonetheless, \steer does have attractive practical attributes and can indeed be an effective full-duplex solution. 
In fact, \steer actually follows our proposed framework in \figref{fig:framework} but relies solely on measurements of self-interference and does not make use of optional downlink, uplink, or cross-link measurements. 
In the next subsection, we extend \steer to incorporate these to overcome its aforementioned practical shortcomings.




\subsection{\steerp: A More Robust Version of \steer}





We now introduce \steerp, a more robust version of \steer that makes use of downlink and uplink measurements to guarantee their quality does not prohibitively degrade when slightly shifting beams to reduce self-interference.
To do this, we modify problem \eqref{eq:steer} as follows by slightly altering the self-interference constraint as \eqref{eq:steerp-target} and introducing a new spectral efficiency constraint \eqref{eq:steerp-se}.
\begin{subequations}\label{eq:steerp}
    \begin{align}
    \min_{\thtx,\thrx} \ \min_{\Delta\varthetatx,\Delta\varthetarx} \ & \Delta\varthetatx^2 + \Delta\varthetarx^2 \\ 
    \st \ 
    & \inrrx\thtxrx \leq \inrrxthresh \label{eq:steerp-target} \\
    & \sesum\thtxrx \geq \minop{\sesumtgt,\sesummax} \label{eq:steerp-se} \\
    &\thtx \in \thtxinit+ \setnbr\parens{\Delta\varthetatx,\delta\thetatx} \\
    &\thrx \in \thrxinit + \setnbr\parens{\Delta\varthetarx,\delta\thetarx} \\
    & 0 \leq \Delta\varthetatx \leq \Delta\thetatx, \ 0 \leq \Delta\varthetarx \leq \Delta\thetarx
    \end{align}
\end{subequations}
Problem \eqref{eq:steerp} aims to find the beam pair $\pthtxrx$ that yields a sum spectral efficiency above some target $\sesumtgt$ while minimizing the beam pair's distance from the initial beam selections $\pthtxrxinit$.
In doing so, it is required that self-interference be below some threshold $\inrrxthresh$ and that the beams come from within their respective neighborhoods.
By including \eqref{eq:steerp-se}, we can ensure that a certain level of quality is maintained on the downlink and uplink for an appropriately chosen target $\sesumtgt$.
To outright maximize sum spectral efficiency, one can set $\sesumtgt = \infty$.
The $\min$ operation in \eqref{eq:steerp-se} is simply to ensure this constraint is feasible, where $\sesummax$ is the maximum sum spectral efficiency possible, given the other constraints.


\begin{algorithm}[!t]
\small
    \begin{algorithmic}[0]
        \REQUIRE $\thtxinit$, $\thrxinit$, $\inrrxthresh$, $\sesumtgt$, $\Delta\thetatx$, $\Delta\thetarx$, $\delta\thetatx$, $\delta\thetarx$ 
        \STATE $\txdirsetmeas = \thtxinit + \setnbr\parens{\Delta\thetatx,\delta\thetatx}$, $\rxdirsetmeas = \thrxinit + \setnbr\parens{\Delta\thetarx,\delta\thetarx}$
        \STATE $\set{D}_{\labeltx} = \braces{\Delta\varthetatx = \bars{\thetatx-\thtxinit}: \thtx \in \txdirsetmeas}$, $\set{D}_{\labelrx} = \braces{\Delta\varthetarx = \bars{\thetarx-\thrxinit}: \thrx \in \rxdirsetmeas}$
        \STATE $\set{D} = \braces{\Delta\varthetatx^2 + \Delta\varthetarx^2 : \Delta\varthetatx \in \set{D}_{\labeltx}, \Delta\varthetarx \in \set{D}_{\labelrx}}$
        \STATE $\brackets{\sim,\set{J}} = \mathrm{sort}\parens{\set{D},\mathsf{ascend}}$
        \STATE Approximate (or measure and feed back) cross-link interference $\inrtx$.
        \STATE $\thtxopt = \thtxinit$, $\thrxopt = \thrxinit$, $\sesummax = 0$
        \FOR{$\pthtxrx \in \entry{\txdirsetmeas \times \rxdirsetmeas}{\set{J}}$}
        \STATE Measure $\inrrx\thtxrx$  (or reference if already measured).
        \IF{$\inrrx\thtxrx \leq \inrrxtgt$}
        \STATE Self-interference threshold met; trigger a downlink and uplink quality check.
        \STATE Measure $\snrtx\pthtx$ and $\snrrx\pthrx$ (or reference if already measured).
        \STATE $\sinrtx\pthtx = \snrtx\pthtx / \parens{1 + \inrtx}$
        \STATE $\sinrrx\pthtxrx = \snrrx\pthrx / \parens{1 + \inrrx\pthtxrx}$
        \STATE $\setx\pthtx = \logtwo{1 + \sinrtx\pthtx}$, $\serx\pthtxrx = \logtwo{1 + \sinrrx\pthtxrx}$
        \IF{$\setx\pthtx + \serx\pthtxrx > \sesummax$}
        \STATE $\thtxopt = \thtx$, $\thrxopt = \thrx$, $\sesummax = \setx\pthtx + \serx\pthtxrx$
        \IF{$\sesummax \geq \sesumtgt$}
        \STATE Break for-loop; sum spectral efficiency target met; no further measurements required.
        \ENDIF
        \ENDIF
        \ENDIF
        \ENDFOR
        \ENSURE $\thtxopt$, $\thrxopt$
    \end{algorithmic}
    \caption{Executing \steerp iteratively while collecting a reduced number of measurements.}
    \label{alg:algorithm-solve}
\end{algorithm}

To better understand the motivation behind \steerp and its design problem \eqref{eq:steerp}, consider the approach to solving problem \eqref{eq:steerp} shown in \algref{alg:algorithm-solve}, akin to \cite[Algorithm~1]{roberts_steer}.
Our algorithm begins by forming spatial neighborhoods around the initial transmit and receive beam selections based on some specified size and resolution.
Then, the beam pairs from these neighborhoods are sorted based on their deviation from the initial beam selections.
Cross-link interference between the uplink and downlink users is either approximated or directly measured and fed back to the \bs.
For each beam pair $\pthtxrx$ within the set of sorted candidate beam pairs, self-interference $\inrrx\pthtxrx$ is measured at the full-duplex \bs.
If the measured self-interference is below the threshold $\inrrxtgt$, this triggers measurement of the downlink and uplink \gpsnr.
Note that if $\snrtx\pthtx$ or $\snrrx\pthrx$ have been measured previously, the prior measurements can be referenced to reduce overhead.
With these, the downlink and uplink \gpsinr can be computed, along with the sum spectral efficiency.
This process continues for each candidate beam pair until a beam pair offers a sum spectral efficiency greater than or equal to the target $\sesumtgt$. 
If this target is never met, the beam pair maximizing spectral efficiency will be used; this is the motivation behind $\sesummax$ in \eqref{eq:steerp-se}.
Note that our algorithm handles the case where problem \eqref{eq:steerp} is infeasible by defaulting to the initial steering directions if no beam pair meets the threshold $\inrrxthresh$.

The role of $\inrrxtgt$ in \steerp is to throttle the consumption of resources used to measure downlink and uplink \gpsnr. 
A very low $\inrrxtgt$ will lead to fewer downlink and uplink measurements (saving on overhead) but may prevent \steerp from locating the beam pair that maximizes spectral efficiency.
Therefore, it is essential that $\inrrxtgt$ not be too strict; letting $\inrrxtgt = \infty$ will trigger measurement of $\snrtx\pthtx$ and $\snrrx\pthrx$ for all $\thtx$ and $\thrx$ until the target $\sesumtgt$ is met---this may be resource-expensive.
As we will see, $\inrrxtgt \approx 6$ dB is sufficient in our experimental evaluation of \steerp. 
In addition, a modest target $\sesumtgt$ can reduce the execution time and number of measurements required by \steerp but can also restrict it from maximizing spectral efficiency; recall $\sesumtgt = \infty$ will force it to maximize spectral efficiency.
As the neighborhoods $\parens{\Delta\thetatx,\Delta\thetarx}$ widen, the sum spectral efficiency cannot degrade with \steerp, unlike with \steer, as we will see shortly.
Keeping the neighborhood small, however, can help keep the number of measurements required low, especially when $\inrrxthresh$ and $\sesumtgt$ are high. 










\begin{figure}
    \centering
    \includegraphics[width=\linewidth,height=0.2\textheight,keepaspectratio]{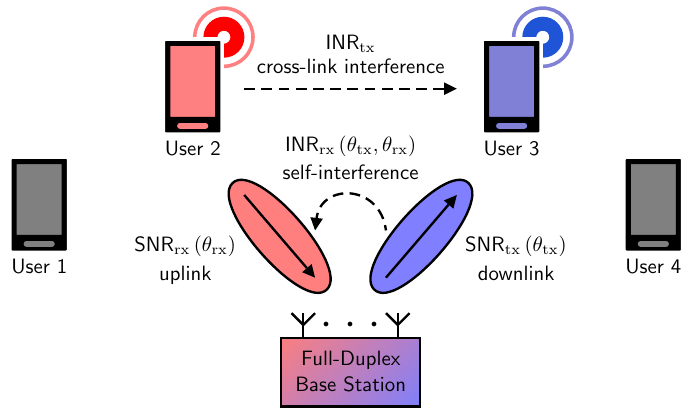}
    \caption{In the lobby shown in \figref{fig:lobby}, we distributed four single-antenna users around our experimental full-duplex \bs. Each user can either transmit uplink or receive downlink at any given time. We implemented \steer and \steerp and experimentally evaluated their performance across all downlink-uplink combinations through measurement of self-interference, downlink \gsnr, uplink \gsnr, and cross-link interference.}
    \label{fig:steer-setup}
    \vspace{-0.5cm}
\end{figure}

\subsection{Real-World Evaluation of \steer and \steerp}

\begin{figure*}
    \centering
    \subfloat[Downlink \gsnr, $\snrtx$.]{\includegraphics[width=\linewidth,height=0.23\textheight,keepaspectratio]{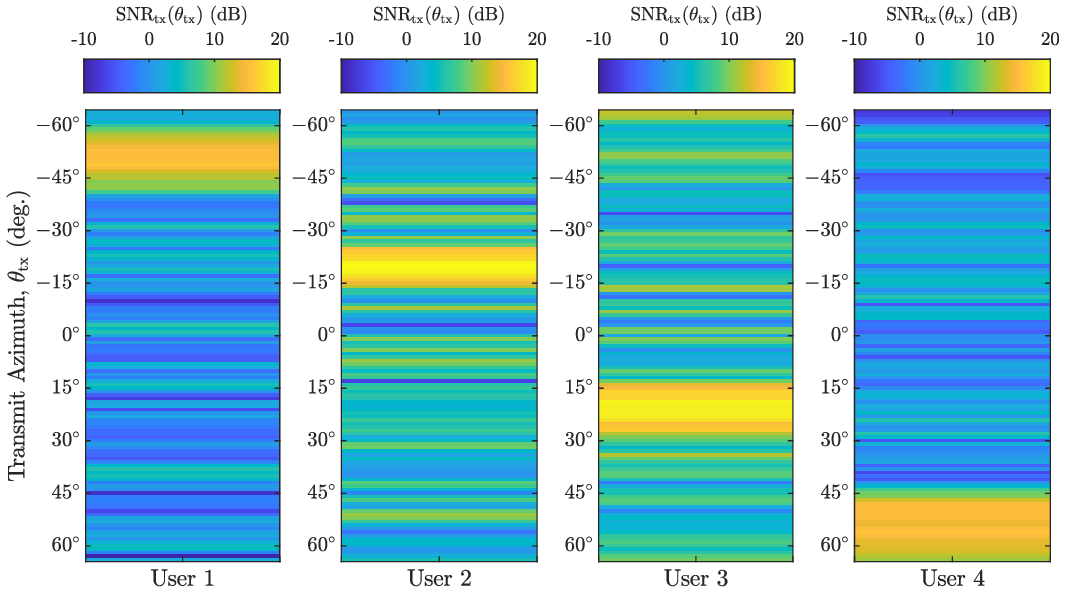}
        \label{fig:snr}}
    \hfill
    \subfloat[Cross-link interference, $\inrtx$.]{\includegraphics[width=\linewidth,height=0.23\textheight,keepaspectratio]{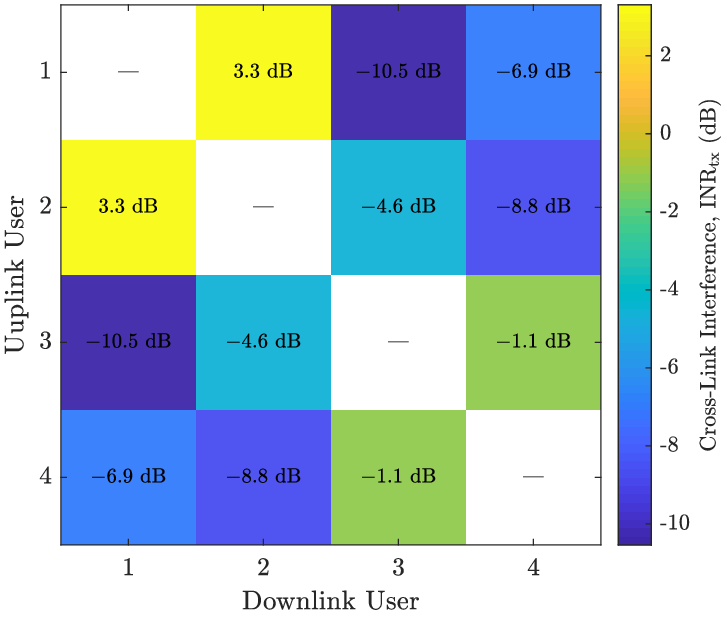}
        \label{fig:inr-tx}}
    \caption{(a) Measured downlink \gsnr for each of the four single-antenna users as a function of transmit steering direction $\thetatx$. (b) Measured cross-link interference $\inrtx$ between each downlink-uplink user pair.}
    \label{fig:snr-inr-tx}
    \vspace{-0.5cm}
\end{figure*}

Now, we conduct an experimental evaluation of \steer and \steerp using our off-the-shelf phased arrays.
As illustrated in \figref{fig:steer-setup}, we distributed four single-antenna users around our experimental full-duplex \bs in the lobby shown in \figref{fig:lobby}.
The transmit and receive arrays at the full-duplex \bs were arranged in the default side-by-side configuration with the transmit array on right and receive array on left, separated by 10 cm.
One of the same 60 GHz arrays was used for each \ue, activating only a single antenna for an omnidirectional pattern.
Each \ue operates in a half-duplex \tdd fashion, meaning the full-duplex \bs is capable of transmitting downlink to any \ue while receiving uplink from any of the other three \ues.
In this experimental setup, we can accurately gauge system performance by measuring self-interference, cross-link interference, downlink \gsnr, and uplink \gsnr. 
In \figref{fig:snr}, we plot downlink \gsnr as a function of \bs steering direction $\thetatx$ for each of the four \ues, which shows them distributed around $-50^\circ$, $-20^\circ$, $20^\circ$, and $50^\circ$ in azimuth from the \bs.
Our measurements of cross-link interference for each downlink-uplink user pair are shown in \figref{fig:inr-tx}, which ranges from about $-10$ dB to $3$ dB and was symmetric between any two \ues, given they are identical. 

For any $\pthtxrx$, we can measure the resulting $\sinrtx\pthtx$ and $\sinrrx\pthtxrx$, which can then be used to directly compute achievable spectral efficiencies 
\begin{align}
\setx\pthtx &= \logtwo{1 + \sinrtx\pthtx}, \qquad
\serx\pthtxrx = \logtwo{1 + \sinrrx\pthtxrx}, 
\end{align}
with their sum being denoted $\sesum\pthtxrx$.
When running \steer and \steerp, we use a spatial resolution of $\delta\thetatx = \delta\thetarx = 1^\circ$, the same as used in \cite{roberts_steer}.
For \steer, we use a self-interference target of $\inrrxtgt = 0$ dB, since this was found to yield best performance.
To demonstrate \steerp, we use a target of $\inrrxtgt = \sesumtgt = \infty$, which will maximize sum spectral efficiency. 
In both, for initial beam selection, we execute exhaustive beam alignment using codebooks spanning $-60^\circ$ to $60^\circ$ with $8^\circ$ resolution, defined as $\sCtx = \sCrx = \braces{-60^\circ,-52^\circ,\dots,60^\circ}$.
When presenting sum spectral efficiency, we normalize it for illustrative purposes as
\begin{align}
\frac{\sesum\parens{\thtx,\thrx}}{\underbrace{\logtwo{1 + \snrtx\parens{\thtxinit}} + \logtwo{1 + \snrrx\parens{\thrxinit}}}_{\textsf{``}\mathsf{codebook~capacity}\textsf{''}}},
\end{align}
which simply normalizes it to the interference-free full-duplex capacity following beam alignment, which we refer to as the \textit{codebook capacity}.
A normalized sum spectral efficiency of $0.5$ can be obtained by half-duplex operation via \tdd, whereas around $1$ is approximately best-case performance to expect from a full-duplex solution following beam alignment.


\begin{figure*}
    \centering
    \subfloat[Sum spectral efficiency.]{\includegraphics[width=\linewidth,height=0.26\textheight,keepaspectratio]{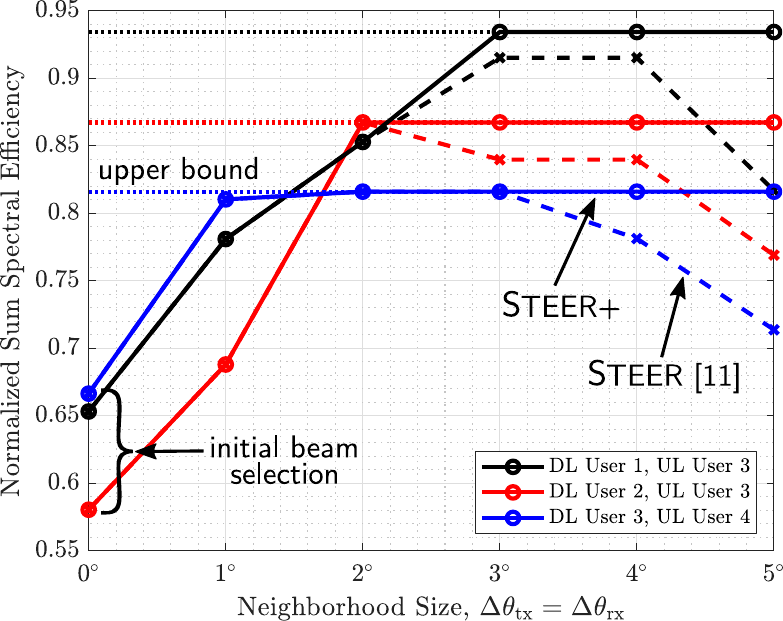}
        \label{fig:nbr-se}}
    \qquad
    \subfloat[Downlink quality, $\sinrtx$.]{\includegraphics[width=\linewidth,height=0.26\textheight,keepaspectratio]{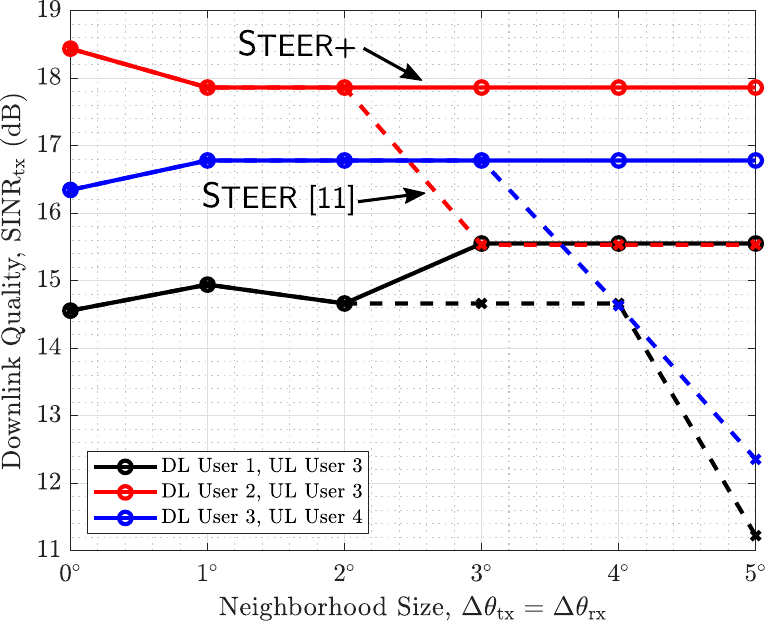}
        \label{fig:nbr-sinr}}
    \caption{For \steer \cite{roberts_steer} (dashed lines) and \steerp (solid lines), shown is the (a) normalized sum spectral efficiency and (b) downlink \gsinr as a function of neighborhood size $\Delta\thetatx = \Delta\thetarx$ for three different downlink-uplink user pairs. \steer can be sensitive to neighborhood size, whereas \steerp only improves as the neighborhood is widened, demonstrating robustness.}
    \label{fig:nbr-se-sinr}
\end{figure*}


In \figref{fig:nbr-se}, for three user pairs, we plot normalized sum spectral efficiency as a function of neighborhood size $\Delta\thetatx = \Delta\thetarx$ for both \steer (dashed lines) and \steerp (solid lines).
The maximum achievable sum spectral efficiency (i.e., $\sesummax$ for $\Delta\thetatx=\Delta\thetarx = \infty^\circ$) for each user pair is shown as the dotted lines.
With the initial beam selections (i.e., when $\Delta\thetatx = \Delta\thetarx = 0^\circ$), all three user pairs marginally exceed $0.5$. 
When activating \steer and \steerp with $\Delta\thetatx = \Delta\thetarx = 2^\circ$, noteworthy increases in spectral efficiency are enjoyed. 
\steer, however, degrades as neighborhood size increases beyond a certain point for all three user pairs, and notice that the point of degradation differs across user pairs.
This degradation is due to the simple fact that \steer does not take into account $\snrtx$ or $\snrrx$ but rather aims to purely minimize self-interference $\inrrx$.
As such, if permitted to deviate too far, \steer may select beam pairs that do not maintain high \gsnr on the downlink and uplink.
This highlights \steer's sensitivity to the selection of neighborhood size (a design parameter).
Choosing $\Delta\thetatx = \Delta\thetarx = 2^\circ$ yields performance better than initial beam selection but is not optimal for all users.
\steerp, on the other hand, can obtain the maximum sum spectral efficiency possible in all three cases with a neighborhood size of $3^\circ$; we found this to be true across all twelve downlink-uplink user pairs.

To better explore the discrepancy between \steer and \steerp, consider \figref{fig:nbr-sinr} showing downlink \gsinr for each of the same three user pairs as in \figref{fig:nbr-se}.
Following beam alignment, downlink \gsinr is expectedly quite high, given the fairly modest cross-link interference levels observed in \figref{fig:inr-tx}.
\steer and \steerp closely follow each other for small neighborhoods but diverge beyond a certain point.
In its effort to purely minimize self-interference, \steer sacrifices downlink \gsnr by selecting a beam pair prohibitively far from the initial beam selection when allocated a larger neighborhood.
\steerp, on the other hand, strategically trades off downlink \gsinr for self-interference reduction only when it improves sum spectral efficiency.
In fact, by doing this, \steerp can actually \textit{improve} downlink \gsinr by refining its beam selection following beam alignment, as seen in the blue and black lines.
Referring to our proposed framework in \figref{fig:framework}, the improvements offered by \steerp over \steer highlight the impact that downlink and uplink measurements can have on a design's performance.
Summarizing concisely, the added robustness of \steerp comes at a cost of extra measurements, which consume radio resources.

\begin{figure*}
    \centering
    \subfloat[Self-interference $\inrrx$ when activating \steerp.]{\includegraphics[width=\linewidth,height=0.26\textheight,keepaspectratio]{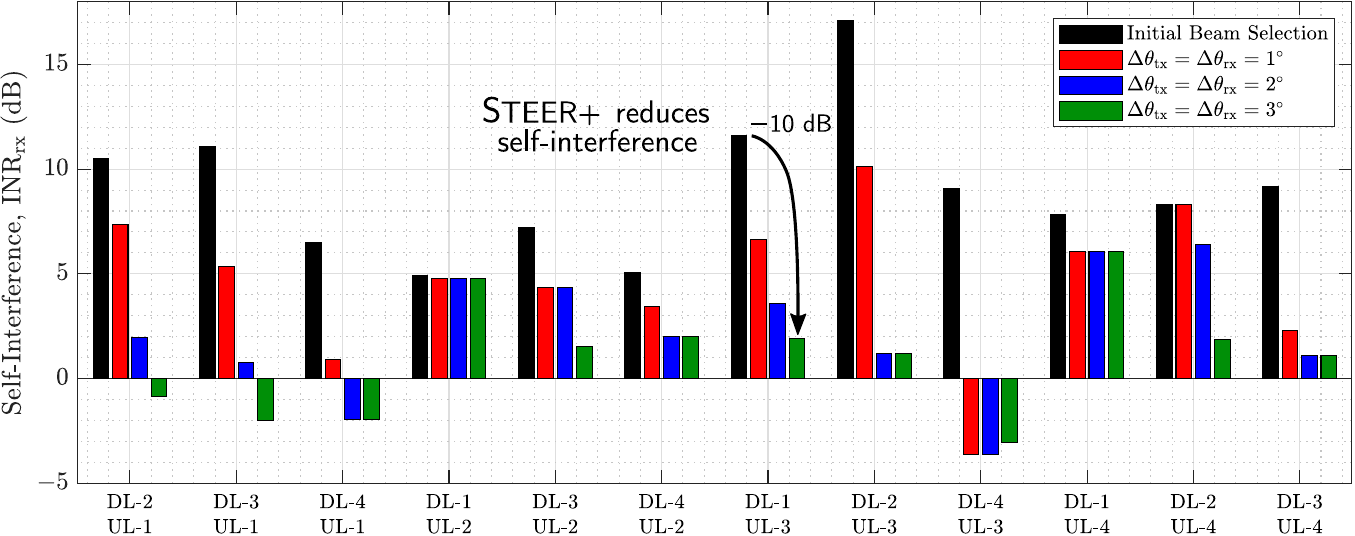}
        \label{fig:bar-inr}}
    \\
    \subfloat[Uplink quality $\sinrrx$ when activating \steerp.]{\includegraphics[width=\linewidth,height=0.26\textheight,keepaspectratio]{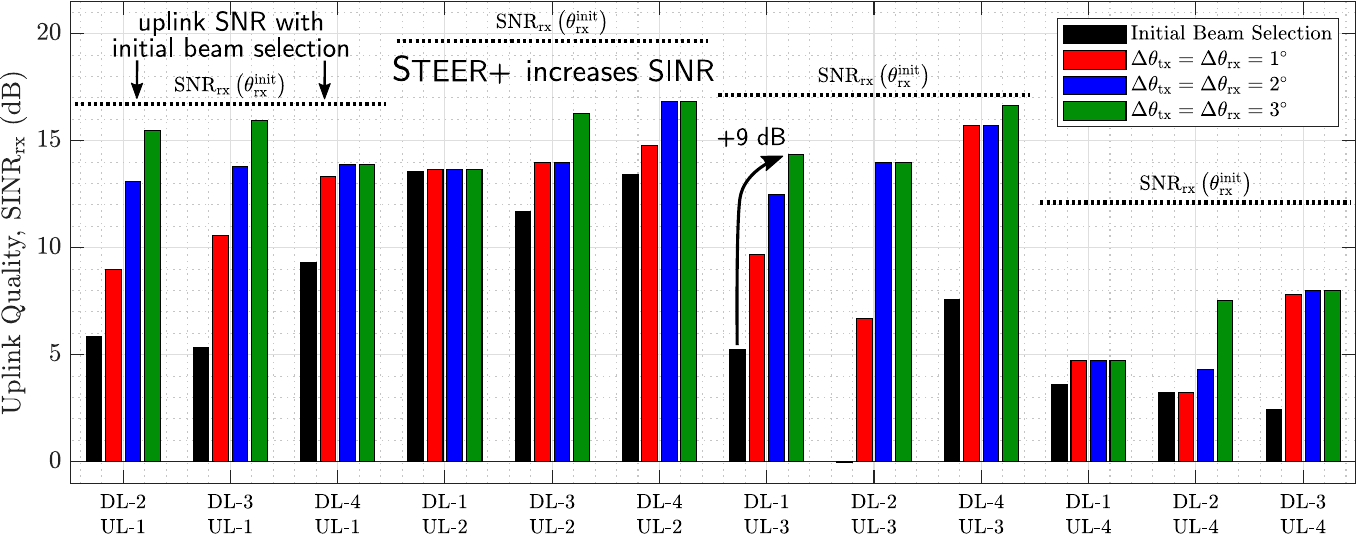}
        \label{fig:bar-sinr}}
    \\
    \subfloat[Sum spectral efficiency when activating \steerp.]{\includegraphics[width=\linewidth,height=0.26\textheight,keepaspectratio]{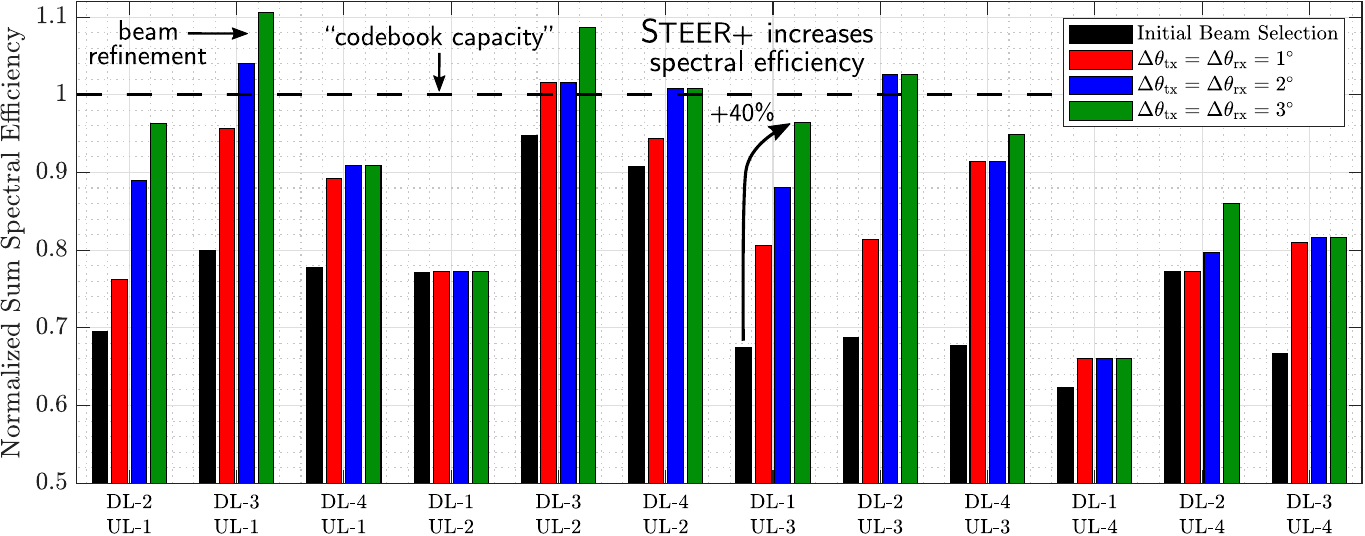}
        \label{fig:bar-se}}
    \caption{The (a) self-interference, (b) uplink \gsinr, and (c) sum spectral efficiency for each downlink-uplink user pair after running \steerp with various neighborhood sizes $\Delta\thetatx = \Delta\thetarx$. In each, the black bars correspond to quantities with initial beam selection, before \steerp is applied. \steerp reduces self-interference, increases \gsinr, and increases spectral efficiency.}
    \label{fig:bar-inr-sinr}
\end{figure*}

To conclude, for completeness, we plot in \figref{fig:bar-inr} the resulting self-interference $\inrrx$ for all twelve downlink-uplink user pairs after running \steerp for various $\Delta\thetatx = \Delta\thetarx$.
In \figref{fig:bar-sinr} and \figref{fig:bar-se}, we analogously plot the resulting uplink \gsinr and sum spectral efficiency, respectively.
In all three, the black bars correspond to quantities with initial beam selection, before \steerp is applied.
\steerp's ability to reduce self-interference is on full display in \figref{fig:bar-inr}.
Without \steerp, self-interference is typically well above the noise floor, ranging from about $5$~dB to $17$ dB above noise.
With \steerp, self-interference can be reduced to just above the noise floor or even below with $\Delta\thetatx = \Delta\thetarx = 3^\circ$.
For $\Delta\thetatx = \Delta\thetarx = 3^\circ$, we can see from \figref{fig:bar-inr} that choosing $\inrrxthresh \approx 6$ dB would not prevent \steerp from maximizing spectral efficiency, allowing it to save on downlink/uplink measurements, since $\inrrx \leq 6$ dB can be achieved by all user pairs.

Reductions in self-interference translate to increases in uplink \gsinr, as seen in \figref{fig:bar-sinr}. 
For nine of the twelve user pairs, \steerp delivers uplink \gpsinr of at least around $13$ dB with $\Delta\thetatx = \Delta\thetarx = 3^\circ$---a transformative improvement in link quality relative to initial beam selection.
In fact, in several cases, \steerp can result in uplink \gpsinr that approach the interference-free \gsnr under conventional beam selection.
These \gsinr increases yield spectral efficiency increases, as shown in \figref{fig:bar-se}.
\steerp fetches at least around $80$\% of the codebook capacity for ten of the twelve user pairs for $\Delta\thetatx = \Delta\thetarx = 3^\circ$.
Notice that some user pairs offer a normalized spectral efficiency exceeding $1$; this reinforces the fact that the codebook capacity is not the true Shannon capacity of the full-duplex system but rather a good baseline for us to compare against under conventional beam selection.
This also highlights how \steerp can be used to not only reduce self-interference but also to refine the beams following initial beam selection to improve \gsnr, further justifying its overhead and potentially allowing for coarser beam alignment. 
Additionally, this shows that \steerp can even outperform conventional beam selection with \textit{ideal} analog/digital self-interference cancellation since it can improve \gsnr in its efforts to reduce self-interference.

\steerp struggles to offer appreciable spectral efficiencies for two user pairs: (DL-1, UL-2) and (DL-1, UL-4).
For these user pairs, \steerp cannot locate very low self-interference without prohibitively degrading downlink and uplink beamforming gain, forcing it to settle for modest \gpsinr.
Referring to the self-interference profile of the lobby in \figref{fig:env-lobby}, we can see that transmitting in the vicinity of $\thetatx \approx -60^\circ$ and receiving around $\thetarx \approx 60^\circ$ yielded highest self-interference. 
This makes transmitting downlink to user 1 while receiving uplink from user 4 particularly challenging, hence \steerp's struggles.
Ultimately, these results highlight the potential for user selection to be powerful tool in improving full-duplex performance when using beamforming-based solutions such as \steerp.
For instance, serving (DL-1, UL-3) in one time slot and then (DL-4, UL-2) in the next would presumably be preferred over serving (DL-1, UL-2) and then (DL-4, UL-3)---an average normalized sum spectral efficiency of 0.95 versus 0.82.
One can imagine that this power of user selection would magnify when given an even larger pool of users to serve, making this a very interesting topic for future work.
Additionally, there is always the potential to apply other methods of self-interference cancellation atop \steerp to fetch even higher spectral efficiencies.

\section{Conclusion and Topics for Future Work} \label{sec:conclusion}


In this work, we conducted a real-world evaluation of full-duplex \mmwave systems using off-the-shelf 60 GHz phased arrays.
We began by presenting self-interference measurements and investigating the effects a variety of factors have on these measurements, including array positioning, the environment, and beam steering direction. 
While our measurements may not extend exactly to other systems and settings beyond our own, the takeaways and conclusions can be used to drive system design and to motivate future measurement and modeling campaigns.
We highlighted five important challenges faced by practical full-duplex \mmwave systems including beam alignment, limited channel knowledge, and adaptability.
To address some of these challenges, we presented a general measurement-driven framework for beamforming-based full-duplex solutions. 
Based on this framework, we introduced \steerp, a novel beamforming-based solution inspired by the recent work of \steer \cite{roberts_steer}.
We experimentally evaluated both \steer and \steerp to demonstrate how they both can be effective solutions, with \steerp's added robustness highlighting the importance of downlink and uplink measurements. 

This work has motivated a variety of future research directions.
Additional measurements of self-interference in a variety of environments and configurations are always welcome contributions to the full-duplex community; in particular, investigating the time-variability of self-interference in real-world settings and the extremely small-scale spatial variability (within one degree) would address open questions.
Development and experimental evaluation of other beamforming-based solutions---potentially based on the proposed framework herein---would be valuable strides toward practical full-duplex \mmwave systems; perhaps machine learning could be a useful tool in this regard.
Finally, it would be valuable to further characterize the network-level enhancements offered by full-duplex \mmwave systems compared to traditional duplexing strategies, especially when taking into account the configuration overhead of full-duplex solutions. 




\section*{Acknowledgments}

This work was supported in part by the National Science Foundation under Grant No. 1610403. 
Any opinions, findings, and conclusions or recommendations expressed in this material are those of the authors and do not necessarily reflect the views of the National Science Foundation.


\bibliographystyle{bibtex/IEEEtran}
\bibliography{bibtex/IEEEabrv,refs}

\end{document}